\documentclass[preprint,showkeys,superscriptaddress,amsmath,amssymb, bibnotes
]{revtex4-2}

\usepackage{graphicx}% Include figure files
\usepackage{dcolumn}% Align table columns on decimal point
\usepackage[font=small]{caption}
\usepackage{bm}% bold math
\usepackage{hyperref}% add hypertext capabilities
\begin{document}

\title{\textbf{Learning to generalize in evolution through annealed population heterogeneity} 
}% 

\author{Federica Ferretti}
 \email{Contact author: fferretti@northwestern.edu}
 \affiliation{NSF-Simons National Institute for Theory and Mathematics in Biology, Northwestern University and University of Chicago, 60611 Chicago, IL.}%Lines break automatically or can be forced with \\
\author{Mehran Kardar}%
 \affiliation{Department of Physics, Massachusetts Institute of Technology, 02139 Cambridge, MA.}%Lines break automatically or can 

\author{Arvind Murugan}
\affiliation{
 Department of Physics, University of Chicago, 60637 Chicago, IL. 
}%

\date{\today}% It is always \today, today,

\begin{abstract}
Evolutionary systems must learn to generalize, often extrapolating from a limited set of selective conditions to anticipate future environmental changes. The mechanisms enabling such generalization remain poorly understood, despite their importance to predict ecological robustness, drug resistance, or design future-proof vaccination strategies. Here, we demonstrate that annealed population heterogeneity, wherein distinct individuals in the population experience different instances of a complex environment over time, can act as a form of implicit regularization and facilitate evolutionary generalization. Mathematically, annealed heterogeneity introduces a variance-weighted demographic noise term that penalizes across-environment fitness variance and effectively rescales the population size, thereby biasing evolution toward generalist solutions. This process is indeed analogous to a variant of the mini-batching strategy employed in stochastic gradient descent, where an effective multiplicative noise produces an inductive bias by triggering noise-induced transitions.
 Through numerical simulations and theoretical analysis we discuss the conditions under which variation in how individuals experience environmental selection can naturally promote evolutionary strategies that generalize across environments and anticipate novel challenges.
\end{abstract}

%Use showkeys class option if keyword
\maketitle

Darwinian evolution operates without foresight or centralized design: it is driven solely by differential survival and reproduction in the environments that populations actually encounter, not by optimization for unseen challenges. Yet evolved populations can sometimes cope with genuinely novel situations, implying that past selection can yield solutions that function beyond prior experience. This observation has motivated a long-standing line of work that frames evolution as a form of learning from examples \cite{Valiant-book,Valiant-evolvability,koonin-multilevel}: populations experience a sequence of environments analogous to training samples, and mutation and selection update the genotype distribution in response. A concrete illustration comes from antibody evolution: during affinity maturation, B-cell lineages encounter a limited panel of antigens, yet the resulting repertoire may bind future variants by leveraging conserved chemical or structural features while ignoring strain-specific decorations. Generalization, in this view, is assessed by performance in a previously unseen context: a population is said to generalize if it achieves high fitness (or binding/neutralization) despite the novelty of the challenge. This analogy sharpens a concrete question: under what conditions does selection on past challenges yield solutions that extrapolate to new ones? In machine learning, success depends on algorithmic choices such as loss functions, regularization, data partitioning, and inductive biases \cite{phys-rev-ML,taxonomy,dropout-2014,LeCun2012-efficient-backprop, curriculum-learning}. In evolving systems, the corresponding levers are population dynamics and biological context: which regimes of mutation and recombination rates, effective population size and structure, and environmental and biological heterogeneity most reliably promote generalization?

Here we analyze when evolutionary population dynamics provide precisely such a bias toward generalization. We study populations in which different individuals simultaneously occupy distinct microenvironments (e.g., spatial niches, host states, physiological contexts), and across generations lineages are reshuffled across these contexts (``annealed'' heterogeneity). In this setting, selection is informed not only by average fitness across the distinct challenges in these microenvironments but also by the variability of fitness across them. We show that population heterogeneity in experiencing selection systematically disfavors genotypes with large across-challenge variance, even when they have moderately higher mean fitness, thereby favoring generalist genotypes with more uniform performance. We map parameter regimes, set by heterogeneity, effective population size, and mutation rate, that bias evolution toward generalist solutions.

We then adopt a learning-theoretic perspective that clarifies our results: population heterogeneity in exposure to selection is analogous to structured mini-batching in machine learning. Each generation offers side-by-side samples from distinct microenvironments, enabling selection to directly ``see'' cross-environment fluctuations. The resulting stochastic dynamics act as an implicit regularizer like mini-batching \cite{gen-gap-large-batch,sclocchi-sgd}, penalizing genotypes with high across-environment variance and biasing the system toward invariant, generalist solutions. Conversely, when there is no microenvironmental structure, training proceeds on aggregated data, removing this variance-based signal and encouraging overfitting to idiosyncrasies of the training set.

Numerous prior works have studied conditions that favor the evolution of generalists, most often through temporal changes in the external environment \cite{Shenshen-Arvind-chirp, Sartori-Leibler,Shenshen-basins,Ivana-fate,Shenshen-Arup-HIV, Rao-Leibler}. In contrast, our macro-environment is static; population heterogeneity alone generates the fluctuations needed to select for invariants. %Further, we exploit the analogy to mini-batching and generalization in statistical learning theory\cite{ } to parameterize the mismatch between training and test distributions and study evolutionary outcomes as a function of this gap.
Beyond explaining when evolution generalizes, our perspective suggests actionable levers, such as controlling within-generation heterogeneity and population structure in laboratory evolution, immunization protocols, or microbial selection schemes, to steer evolving systems toward invariants. More broadly, our work points to a design principle: by shaping how complex environments are experienced across a population, we can make evolution prepare that population for novel unseen challenges. 

\subsection*{Model}

We begin with a linear classification toy problem to isolate the core difficulty of generalization and to motivate the structure of the fitness landscapes used later. Consider inputs $\mathbf{x}=(x_0,x_1,x_2,x_3)\in\mathbb{R}^4$ with four features, and consider unit-norm linear classifiers $\mathbf{g}$ with score $s(\mathbf{x})=\mathbf{g}\cdot\mathbf{x}$. 
Training examples arrive from two domains $D_1,D_2$ with domain-specific means $\mu_1=\mathbb{E}[\mathbf{x}\mid D_1]=(2,5,0,0)$ and $\mu_2=\mathbb{E}[\mathbf{x}\mid D_2]=(2,0,5,0)$, while the held-out test domain $D_3$ has $\mu_3=\mathbb{E}[\mathbf{x}\mid D_3]=(2,0,0,5)$.  That is, feature $x_0$ is weak but invariant across domains; features $x_1,x_2,x_3$ are stronger but domain-specific. If we pool $D_1$ and $D_2$ before training, the pooled mean is $\bar{\boldsymbol{\mu}}=(2,2.5,2.5,0)$. Training on the pooled data therefore favors $\mathbf{g}$ aligned with $(0,1,1,0)$ and largely ignores $x_0$. Such a predictor performs well on $D_1,D_2$ but transfers poorly to $D_3$, where the signal has moved to $x_3$. By contrast, a ``generalist'' predictor aligned with $x_0$, e.g. $\mathbf{g}=(1,0,0,0)$, attains a smaller average margin on the training domains but a nonzero margin in every domain, because its signal is invariant; its across-domain performance is flatter (lower mean, much lower variance).

How can a learning algorithm be biased toward such invariant features when pooled training data provides no preference for $x_0$?  This question does not fit into standard learning frameworks that assume that test data come from the same distribution as training data (empirical risk minimization) \cite{VC-nature-theory}. Instead, this setup, where the generalist feature has lower mean performance than specialists in the context of training data, has been studied in learning theory as out-of-domain generalization or as invariant risk minimization \cite{IRM-meta-2019}.  One established option is to \emph{explicitly} regularize, e.g., by augmenting the loss function with a penalty on the variation of domain-wise losses \cite{variance-penalty, entropy-sgd}.  
However, a more intriguing possibility for evolutionary dynamics is an \emph{implicit} regularization or bias that arises from using only a subset of the data at a time. Such techniques, called mini-batching, refer to drawing small, randomly chosen groups of training examples, instead of the whole dataset, so that each update of the algorithm is based on just this small slice. When stochastic gradient updates use mini-batches drawn from a single domain at a time and alternate domains, the resulting gradient fluctuations push toward parameters that equalize domain-wise performance, selecting a classifier $\mathbf{g}$ that depends on invariant features. In contrast, drawing mini-batches from the pooled data $D_1 \cup D_2$ hides cross-domain variation, removes the variance-based signal, and encourages overfitting to domain-specific idiosyncrasies. 

\begin{figure}
    \centering
    \includegraphics[width=\textwidth]{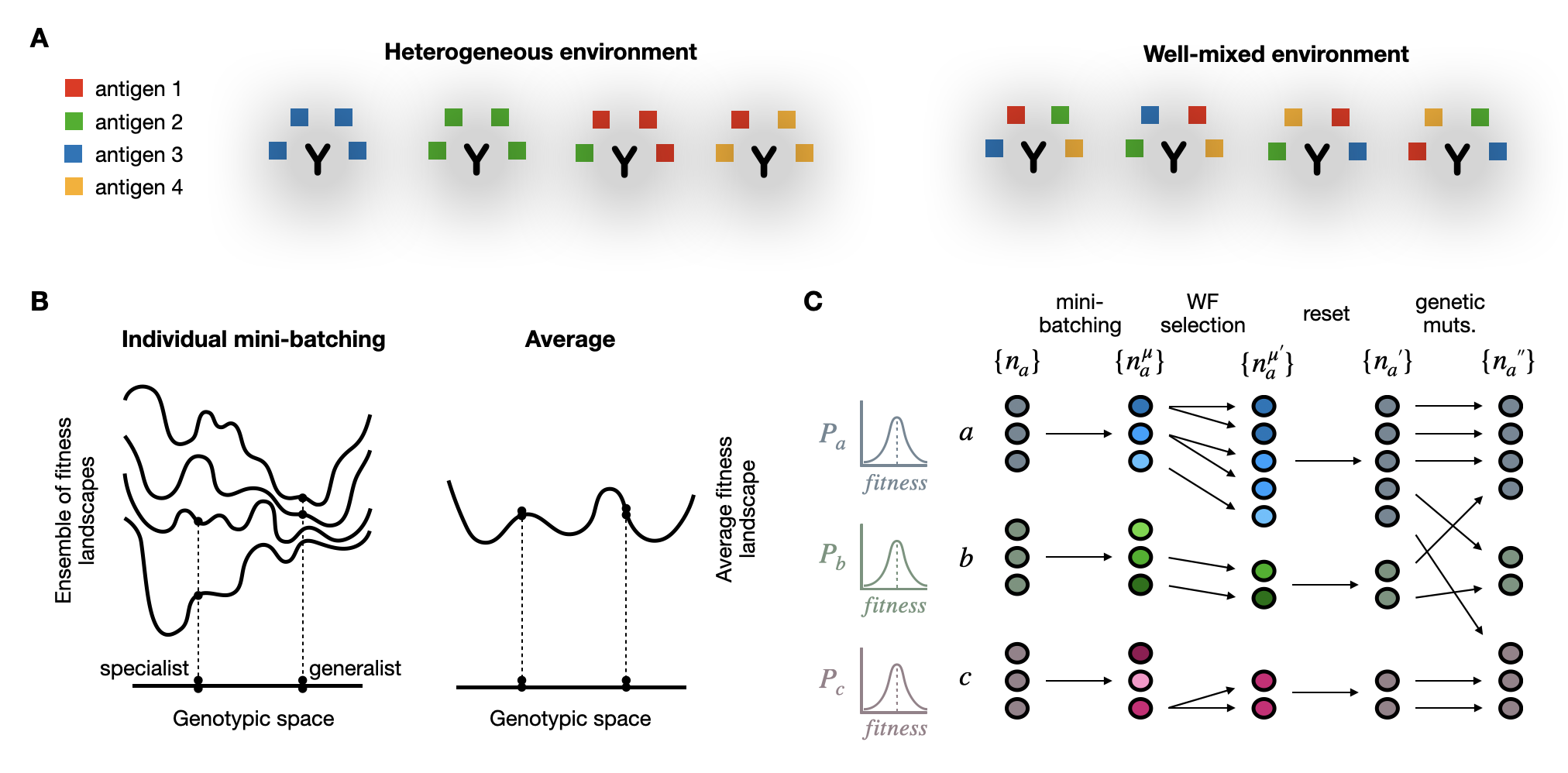}
    \caption{\textbf{Schematic of the evolutionary mini-batch analogy and modeling framework.} \textbf{A Biological inspiration.} (left) Individuals in the population (e.g., antibodies, shown as Ys) encounter distinct microenvironments (e.g., different concentrations of distinct antigens) so that the fitness of every individual is continually reshuffled across generations through exposure to a variety of environmental challenges. In contrast, the fitness landscape is static (right) if the population is evolve din a single homogeneous setting. \textbf{B Fitness landscape model.} %The collection of possible fitness values in all microenvironments form a fitness landscape ensemble. The random selection of a microenvironment corresponds to a mini-batching protocol. The reference process that we compare our results to is evolution in a single, fixed landscape, corresponding to the average fitness landscape. 
    This source of heterogeneity is represented as an ensemble of fitness landscapes; in each generation every individual samples one landscape at random, analogous to a mini-batch of training data in stochastic gradient descent.
    \textbf{C Evolutionary dynamics.} In our model, a single generation consists of four steps: (1) microenvironment sampling by individuals, (2) Wright–Fisher reproduction and selection on the sampled fitnesses, (3) resetting of phenotypes so that microenvironmental states do not persist, and (4) mutation. Together these steps capture how annealed population heterogeneity introduces an additional source of demographic noise that biases evolution toward genotypes with robust, across-environment performance. }
    \label{fig:1}
\end{figure}

Inspired by this analogy, we model an evolving population in a strongly heterogeneous environment where, in every generation, different individuals encounter distinct microenvironments. The subpopulation exposed to any one microenvironment at that time effectively plays the role of a \emph{mini-batch}, providing a particular sample of environmental conditions that drives selection in that generation.
By annealed population heterogeneity we mean that, in each generation, every individual independently resamples its microenvironment from a fixed distribution; microenvironmental states do not persist across generations (i.e., no temporal correlations), nor are coherent across the population, distinguishing this setting from switching ‘seascapes’ \cite{Mustonen-Lassig-PRL2008, Mustonen-Lassig-seascape-review}.

Formally, every individual in the population is characterized by its genotype $\mathbf g_a$ and by the environmental challenge it is exposed to, which we denote by a multidimensional random variable $\mathbf{x}$, where the length of $\mathbf x$ is the number of independent features needed to describe the environment. The fitness of any genotype $\mathbf g_a$ in any environment $\mathbf x$ is a function $f(\mathbf g_a,\mathbf x)$. The key idea we want to incorporate into our model is that certain of these features $\mathbf{x}$ are strongly variable, while others are pretty conserved across microenvironments. As a result, ``generalist'' genotypes that ``align" to the conserved features of $\mathbf{x}$ will have a reliable, almost constant fitness in every condition, while the specialists, which we can think of as genotypes aligning to the variable features, will perform unreliably across microenvironments, even though they might have better fitness on the pooled/homogenized version of this environment (i.e., the average macro-environment). 

Since in our setting $\mathbf x$ is randomly sampled by each individual from a collection of possible micro-environments (the training dataset), we can focus directly on the induced probabilistic genotype-to-fitness map, thus elevating the common picture of a static, average fitness landscape to an ensemble of fitness landscapes (Fig.~\ref{fig:1} B). This ensemble is characterized by the collection of fitness distributions $P_a(f)$ with mean $f_a$ (measuring the average performance of the genotype $\mathbf g_a$ across environments) and variance $V_a$. To simplify the notation and the derivation of the analytical results, we consider discrete distributions of fitness values $\{f_a^\mu\}_{\mu=1,\dots, M}$ which occur with probabilities $\{\mathcal P_a^\mu\}_{\mu=1,\dots, M}$, but all the results still hold in the continuous case.

To summarize, the evolutionary process we aim to study is 
% The full process is described as 
a composite Markov chain, whose transition probability is given by the convolution of the following four sub-steps (illustrated in Fig.~\ref{fig:1} C):
\begin{enumerate}
    \item Mini-batching: At each generation, each member of the population independently picks one of the $M$ fitness values on hand, according to the probability distribution associated to its genotype.
    This can be seen as the mini-batching step thanks to which the annealed population heterogeneity is implemented, since every isogenic subpopulation $\{n_a\}$ is partitioned into phenotypically distinct subpopulations $\{n_a^\mu\}$, according to a multinomial probability distribution:
    \begin{equation}
        B\left(\{n_a^\mu\}_{a=1,\dots, S;\mu=1,\dots, M}\vert \{n_a\}\right) = \prod_{a=1}^S n_a!\prod_{\mu=1}^M\frac{\left(\mathcal P_a^\mu\right)^{n_a^\mu}}{n_a^\mu!}.
    \end{equation}
    \item Wright-Fisher replication and selection: The population is evolved from a state $\{n_a^\mu\}$ to $\{{n_a^\mu}'\}$ via multinomial sampling, with $N$ trials and probabilities of success $f_a^\mu n_a^\mu/\sum_{b,\nu}f_b^\nu n_b^\nu$ for each of the $S\times M$ phenotypic categories indexed by $(a,\mu)$.
    \item Resetting of the phenotype: Assuming no persistence in mini-batching across generations, we ``reset'' the state of the population to $\{n_a'\}$, where $n_a' = \sum_\mu{{n_a^\mu}'}$ for any $a$.
    \item Genetic mutations: Upon replication, mutations can be accumulated with given rates. From these rates, it is possible to define a mutational graph with Laplacian $\nu \Lambda_{ab}$, describing the probability of transitioning between any pair of genotypes $a$ and $b$. 
\end{enumerate}
While the propagator of the composite process cannot be written in an insightful closed form \cite{SI}, a useful characterization of the extra source of stochasticity introduced by the mini-batching procedure is obtained by working in the diffusion approximation limit \cite{Gardiner}, which is discussed in the next section.

\section*{Results}
\subsection*{Characterization of the effective noise}
In the limit of large populations under nearly neutral selection and weak mutation, evolutionary population dynamics can be approximated by a diffusion process. As derived in SI~\cite{SI}, the resulting It\^o stochastic differential equation (SDE) for the fraction of the population $z_a$ in genotype $a$ is,
\begin{equation}
    \partial_t z_a = F_a(\boldsymbol z) + \sigma_{ab}(\boldsymbol z) \xi_b , \qquad \langle\xi_a(t)\xi_b(t')\rangle = \delta_{ab}\delta(t-t'),
    \label{eq:SDE}
\end{equation}
where $z_a(t) = n_a(t)/N$. The drift and diffusion terms of the Ito-SDE read
\begin{flalign}
    \label{drift} F_a(\boldsymbol z) &= F^0_a(\boldsymbol z) - \frac{V_a - \bar V(\boldsymbol z)}{\bar f(\boldsymbol z)^2}z_a,\qquad F_a^0(\boldsymbol z)= N\frac{f_a - \bar f(\boldsymbol z)}{\bar f(\boldsymbol z)}z_a+N\nu\Lambda_{ab}\frac{f_b}{\bar f(\boldsymbol z)}z_b,\\
    \label{diff} D_{ab}(\boldsymbol z) &= \sigma^2_{ab} =  \underbrace{D^0_{ab}(\boldsymbol z)\left(1+\frac{\bar V(\boldsymbol z)}{\bar f(\boldsymbol z)^2}\right)}_{\hat D_{ab}(\boldsymbol z)} + \underbrace{\delta_{ab}z_a\frac{V_a- \bar V(\boldsymbol z)}{\bar f(\boldsymbol z)^2} - z_a z_b \frac{V_a + V_b - 2\bar V(\boldsymbol z)}{\bar f(\boldsymbol z)^2}}_{\Delta D_{ab}(\boldsymbol z)},\\ 
    &\qquad\qquad\qquad\qquad\qquad\qquad\quad D^0_{ab}(\boldsymbol z) = z_a\delta_{ab}-z_az_b,
\end{flalign}
where $\bar f(\boldsymbol z)$ and $\bar V(\boldsymbol z)$ are population averages of the two genotype-dependent observables $f_a$ (fitness mean) and $V_a$ (fitness variance).
The functions $F^0_a(\boldsymbol z)$ and $D^0_{ab}(\boldsymbol z)$ indicate, respectively, the drift vector and the diffusion tensor of the reference diffusion process, describing the evolution of a large population in a fixed fitness landscape, where the fitness value of each genotype $a$ corresponds to the mean $f_a$ (see \ref{fig:1}).

The equations above apply to any system satisfying the diffusion approximation assumptions, i.e. low mean fitness diversity in the population and small mutation rates, without explicit constraints on the relative strength of mutation and selection \cite{SI}. If, additionally, the system is in an evolutionary regime of strong selection and weak mutation, which favors localized population states, the anisotropic part of the correction to the diffusion tensor $\Delta D_{ab}(\boldsymbol z)$ can be neglected, as well as the correction to the drift. Mini-batching effects then reduce to a simple rescaling of the effective population size: 
\begin{equation}
    N_{\mathrm{eff}} \;\approx\;\frac{N}{\,1+\;\bar V(\boldsymbol z)/\bar f(\boldsymbol z)^2},
\end{equation} 
so higher across-environment variance in fitness effectively shrinks population size and favors genotypes with lower $V_a$. As shown below, this simplification captures the main qualitative behaviors of the system. An analogous rescaling arises in the equilibrium regime of population dynamics \cite{Sella2005}, in which mutations are rare and evolution progresses through subsequent fixation or extinction events \cite{Kimura-original,SI}.

As a particular case, complementary to the scenario of interest for generalization, one can consider evolving populations in a genotypic space with constant $V_a$ ---i.e. with uniform phenotypic variance--- or exhibiting an inverse tradeoff between mean and variance on the right front of the convex hull describing the accessible simplex in the $(f,V)$ plane (cfr. Fig.~\ref{fig:2} D) ---i.e. when higher fitness genotypes also have lower fitness variance across microenvironments. In this case, the prediction of the effective Ito-SDE is that annealed population heterogeneity would help the selection of the genotypes that are the fittest in the average macroenvironment when the mutational load is high.

\subsection*{Non-perturbative effects from perturbative inductive bias} 

\begin{figure}
    \includegraphics[width=0.95\linewidth]{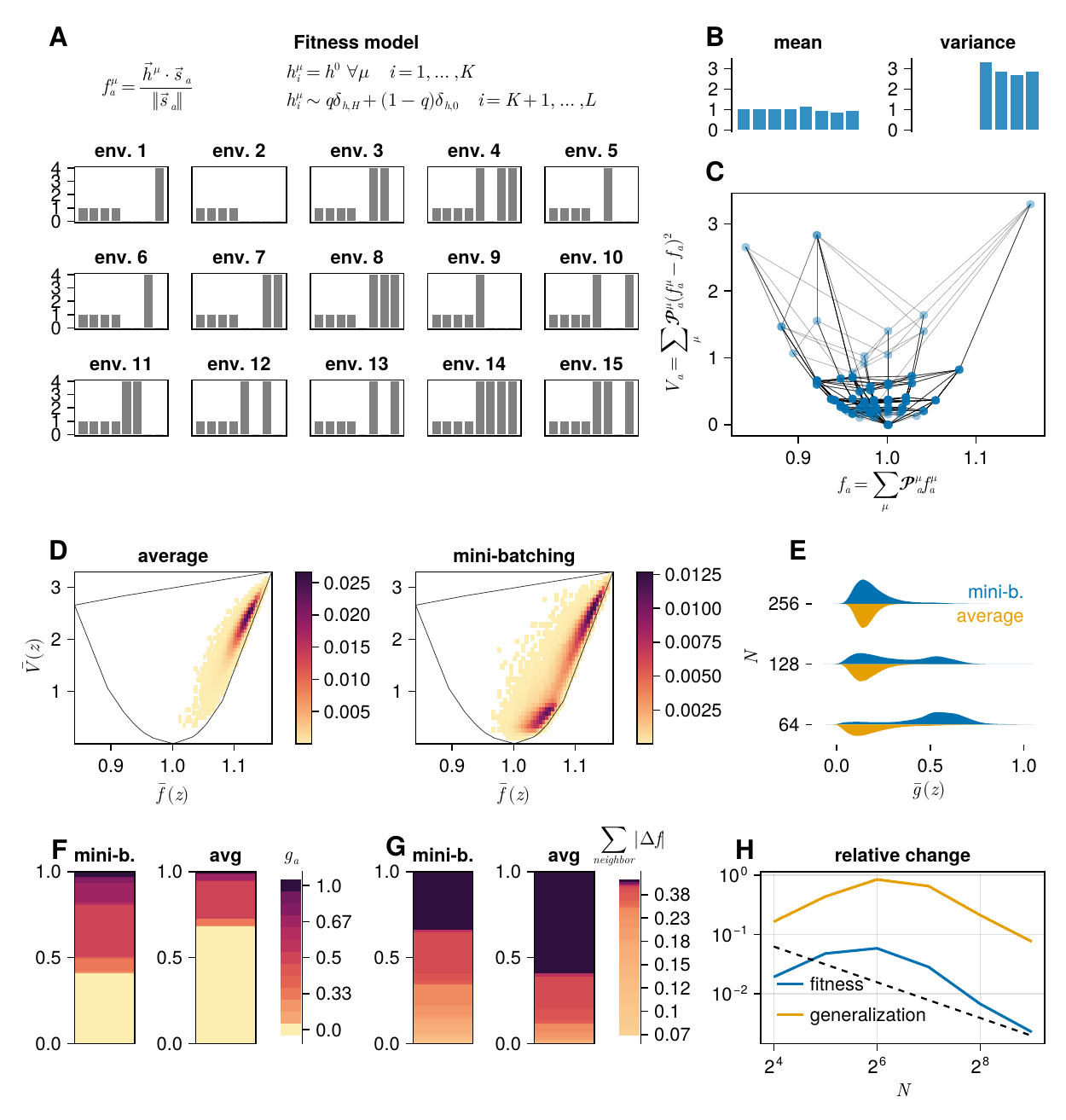}
    \caption{\textbf{Population response to mini-batching in a toy model.} \textbf{A Landscapes.} We generate an ensemble of $M=100$ fitness landscapes with the described probabilistic model, where $q=0.2$, $H=4$, $h^0=1$. At each generation, every individual in the population picks one of these 100 fitness landscapes at random. Genotypic sequences have length $L=8$; selection fields are kept fixed on $K=4$ sites; 15 distinct landscapes were generated (present in the training dataset with different frequency). \textbf{B Statistics of the training set.} Mean and variance of the selection fields across the training dataset.
    \textbf{C Sequence space representation.} The dynamics of the system is governed by the mean ($f_a$) and the variance ($V_a$) of the fitness values of each genotype across the training dataset. The numerical model we considered exhibits a constitutive 
    }
    \label{fig:2}
\end{figure}

\begin{figure}[h]
        \ContinuedFloat
        \caption{ inverse trade-off between these two quantities, as shown by the front on the right end of the plot. In the absence of this trade-off, the fittest sequences would already be the best generalizers, making single-cell variability unnecessary to improve the generalization performance of the population. In the figure, edges connect genotypes that differ for a single mutation.
    \textbf{D Comparison of population dynamics models.} We numerically estimate the joint probability densities of two collective observables, i.e. the population mean fitness $\bar f(z) = \sum f_a z_a$ and the population fitness variance $\bar V(z) = \sum_a V_a z_a$, both for the process with individual mini-batching and for the reference process where the population evolves in the fixed, average landscape. The density is unimodal in the reference case, with a single peak on the top right corner of the simplex, while it is bimodal in the case with individual mini-batching, with a second peak in a lower position on the trade-off front. Population size: $N=128$.
    \textbf{E Distributions at mesoscopic population sizes.} As a function of $N$, the number and location of the modes of the distribution of the collective variable $ \bar g(\boldsymbol z)$ change. At mesoscopic $N$, the difference between the processes with and without mini-batching is most evident.
    \textbf{F Generalization.} We attribute to each sequence a generalization score $g_a$, bounded between 0 and 1, as defined in the main text. The expected population composition by generalization score is significantly different between the reference case (right, dominated by specialists) and the case with mini-batching (left, enriched in generalists). \textbf{G Mutational robustness.} In the presence of mini-batching, the (average) population composition is enriched with "flatter" species, characterized by lower values of the average mutational effects $\gamma_a = \sum_{b} |\Lambda_{ab}(f_b - f_a)|$. 
    \textbf{H Magnitude of mini-batching effect.} We plot the relative changes $2\lvert\langle A\rangle - \langle A\rangle_0\rvert/(\langle A\rangle + \langle A\rangle_0)$ for $A=\bar f(\boldsymbol z)$ and $A=\bar g(\boldsymbol z)$. Here $\langle\cdot\rangle$ and $\langle\cdot\rangle_0$ indicate respectively the average over the steady-state distribution of the process with mini-batching and the reference process. The dashed black line corresponds to $1/N$ (expected order of magnitude of the mini-batching effect). A big deviation from this reference is observed at intermediate population sizes. }
    \end{figure}

Independent sampling of the environment by individuals in the population implies that the impact of mini-batching is an $O(1/N)$ correction to the reference SDE (see \eqref{drift} \eqref{diff}); such an inductive bias would be negligible already at $N \sim 10^2$. Yet simulations on toy fitness landscapes show the opposite: at mesoscopic $N$, the bias reshapes the stationary distribution and shifts population-level observables in a macroscopically detectable way.

Figure \ref{fig:2} shows an example from a toy fitness landscape model on genotypes $\vec s_a\in\{0,1\}^L$. In this model, each genotype is assigned a set of fitness values $f_a^\mu=(\vec h^\mu\cdot \vec s_a)/\lVert \vec s_a\rVert_1$, where $\vec h^\mu$ are random fields with constant first $K$ elements (modeling the conserved environmental features) and independently drawn binary variables on the last $L-K$ elements (variable environmental features).
This model choice yields a natural generalist-specialist axis, quantified by the generalist score
$g_a = (\vec h^0\cdot \vec s_a)/\lVert\vec s_a\rVert_1$, where $\vec h^0$ is the constant field vector, with first $K$ coordinates equal to one and last $L-K$ coordinates equal to zero\footnote{A small uniform field $\epsilon\ll 1$ is added to $\vec h^0$, in order to avoid zero-fitness normalization issues in the Wright-Fisher simulations.}.
Crucially, these choices keep relative fitness differences small (supporting the diffusion approximation) while inducing strong heterogeneity in fitness variance across genotypes. 

At mesoscopic population sizes (i.e. $N\sim 10^2$ in Fig.~\ref{fig:2}), population dynamics with annealed heterogeneity differ markedly from evolution in the averaged fitness landscape. The population composition is more mixed, and enriched in individuals with higher generalist scores (\ref{fig:2} E), or lower fitness variance (the two quantities are inversely related by construction, Fig. S5).
As a result, the expected value of the population-averaged generalist score, $\bar g(\boldsymbol z) = \sum_a g_a z_a$, is higher, while the expected values of the population-averaged fitness variance $\bar V(\boldsymbol{z})$ and mean $\bar f(\boldsymbol z)$ are lower. The steady-state distributions of these collective observables show altered shapes, with the emergence of a new mode at reduced variance states, as visible in Fig \ref{fig:2} D. This behavior can be rationalized as a noise-induced phase transition \cite{transitions-book}, as better explained in the next section. 

Finally, we note that annealed population heterogeneity promotes ``flatter'' genotypes, identified by lower values of the inverse flatness parameter $\gamma_a = \sum_b\vert\Lambda_{ab}(f_a-f_b)\vert$ in Fig \ref{fig:2} G. 
This quantity measures the neutrality of mutations accessible from genotype $a$, in the average landscape. 
The enrichment of lower $\gamma_a$ genotypes mirrors the tendency of stochastic gradient descent to settle in flatter minima \cite{hessian-sgd, TuSGD-PNAS, TuSGD-PRL, tilting-playing} and also echoes the ``survival of the flattest'' principle in evolutionary biology, where lower but flatter fitness peaks are preferred over sharper but higher peaks at high mutation rates \cite{eigen1989molecular,wilke2001flattest}.
Notably, \eqref{drift}--\eqref{diff} do not contain any explicit flatness term. 
Instead, localization into more neutral genotypes arises in our model because flat regions of the average landscape correspond to sub-networks of genotypes with minimal variance in fitness, and population dynamics with mini-batching naturally drive the population towards such genotypes. More generally, we expect that when the fitness variance is smooth across genotypic space, moderate mutation rates, combined with annealed population heterogeneity, will generate a positive feedback that amplifies localization within low-variance, flat-on-average regions.

\subsection*{A non-equilibrium trap is the origin of secondary modes}
The dynamical system described by \eqref{eq:SDE} is generally high-dimensional and its behavior is hard to predict accurately. However, from the form of the drift and diffusion terms in \eqref{drift} and \eqref{diff}, one can deduce that the dynamics must be mostly controlled by the two collective variables $\bar f(\boldsymbol z)=\sum_af_az_a$ and $\bar V(\boldsymbol z)=\sum_aV_az_a$. 
We can derive from \eqref{eq:SDE} the coupled SDEs describing their evolution, which can be read as stochastic extensions of the Price equation:
\begin{eqnarray}
    \label{Price-f}\partial_t \bar f = N \frac{Var_z(f) + \nu\boldsymbol f^\top\Lambda\tilde{\boldsymbol z}}{\bar f} - \frac{Cov_z(f,V)}{\bar f^2} + \eta_f;\\
    \label{Price-V}\partial_t \bar V = N \frac{Cov_z(f,V) + \nu\boldsymbol V^\top\Lambda\tilde{\boldsymbol z}}{\bar f} - \frac{Var_z(V)}{\bar f^2} + \eta_V;
\end{eqnarray}
where $\tilde z_a = f_a z_a$ and $\eta_{f,V}$ are correlated white noises with covariance matrix (see \cite{SI} for full expression):
\begin{equation}
    \langle\eta_f\eta_f\rangle  =\boldsymbol f^\top D(\boldsymbol z) \boldsymbol f=T_f ,\quad  
    \langle \eta_V\eta_V\rangle  =\boldsymbol V^\top D(\boldsymbol z) \boldsymbol V=T_V ,\quad  
    \langle \eta_f\eta_V \rangle =\boldsymbol f^\top D(\boldsymbol z) \boldsymbol V.
\end{equation}
$Var_z(A)= \overline{A^2}- {\bar A}^2$ and $Cov_z(A,B)=\overline{AB}-\bar A\bar B$ represent the variance across the population of observable $A$, and the covariance across the population of $A$ and $B$, respectively.

Although \eqref{Price-f} and \eqref{Price-V} do not form a closed system of equations when the number of genotypes is greater than three, they offer valuable insight into the qualitative behavior of the dynamics in this low-dimensional space.
The deterministic dynamics typically exhibits a separation of scales: first, the system experiences a fast rightward push dictated by the leading order term in \eqref{Price-f}, $N Var_z(f)/\bar f$, which selects populations with homogeneously high fitness. This fast dynamics is then followed by a slow relaxation along the vertical axis towards the mutation-selection-balance fixed point, assuming that high fitness states are degenerate and $\nu\ll1$. For an explicit visualization of the phase portrait in the $(\bar f,\bar V)$ plane, see the discussion of the three-species case in the SI \cite{SI}. 

\begin{figure}
    \centering
    \includegraphics[width=0.95\textwidth]{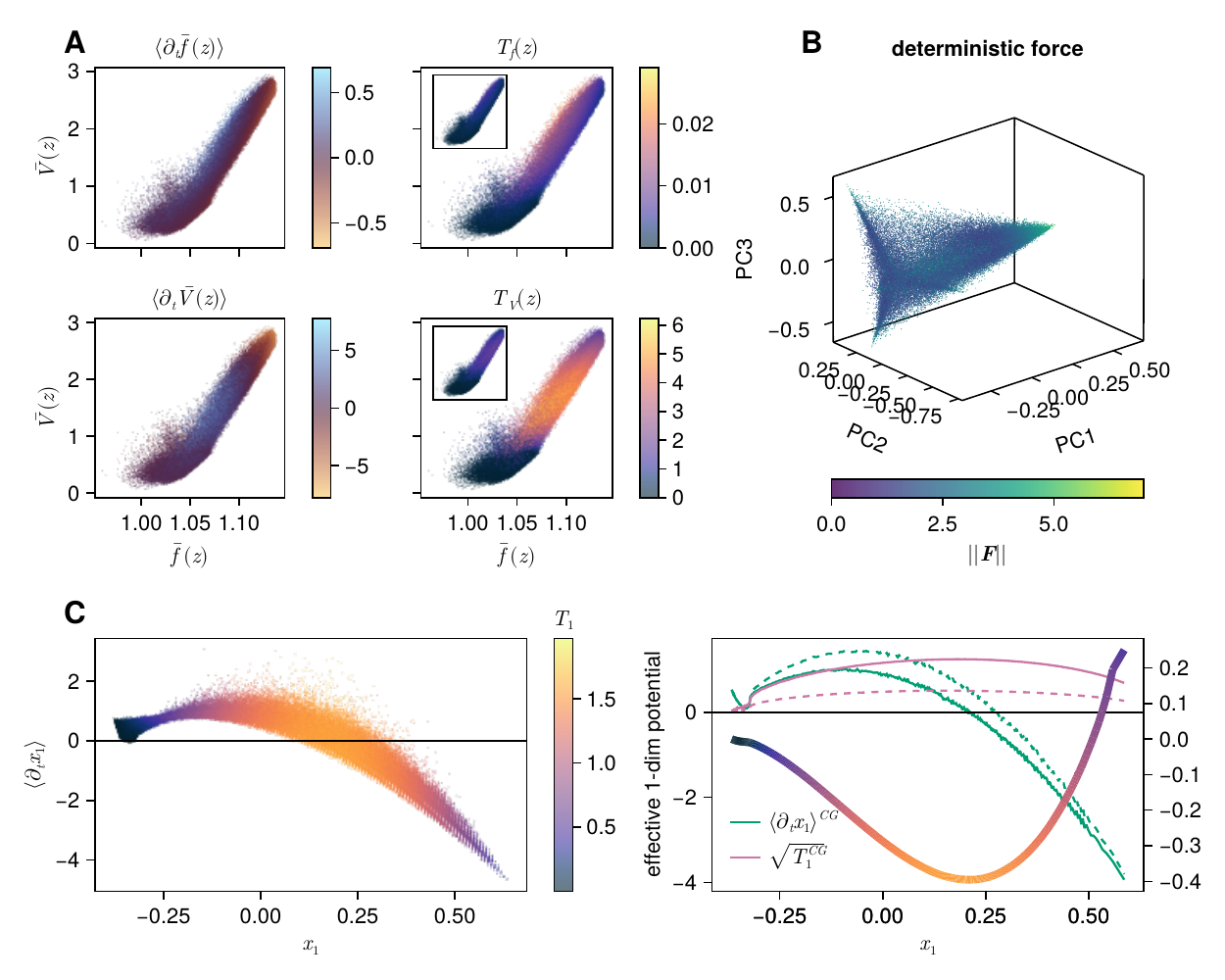}
    \caption{\textbf{Price equation reveals a non-equilibrium trap shaped by demographic noise.} \textbf{A Nullclines and noise amplitude of the stochastic Price equation.} We compute the drift and diffusion terms of the generalized Price equation in \eqref{Price-f}--\eqref{Price-V} across the population states visited at stationarity. The color schemes show the deterministic forces (left) or the temperature of the fluctuations  (right) along the $\bar f$ (top) and $\bar V$ (bottom) axis. The system exhibits multiple points with near-zero drift, and a strongly inhomogenous temperature profile across them. The temperature gradient in the system with mini-batching is an order of magnitude larger than the temperature gradient in the absence of mini-batching (insets).
    \textbf{B Quasi-equilibrium states.} There are two major regions (in PCA representation) where the magnitude of the drift term is near zero. The population may spend long time in these regions, even if not stable fixed points of the dynamics, especially in the presence of noise. 
    \textbf{C Dynamics along PC1.} The projection of the dynamics along the first PC of the ensemble of non-equilibrium steady states at $N=128$ is described by the Price equation in Eq. 23 of the SI ---with $\boldsymbol u$ the eigenvector associated to PC1, such that $x_1=\boldsymbol u \cdot \boldsymbol z=\bar u(\boldsymbol z)$. 
    The system behaves as a quasi-1D dynamical system, as testified by}
    \label{fig:3}
\end{figure}

\begin{figure}[h]
        \ContinuedFloat
        \caption{ the low scatter of the phase portrait in the left panel, both in its deterministic ($y$ coordinate) and stochastic (color) components. 
    On the right, an effective 1D phase portrait is built by binning the PC1 coordinate and averaging over population states in the same bin. The color-changing line is an illustration of the effective pseudo-potential obtained by numerically integrating $\langle\partial_t x_1\rangle^{CG}$. The dashed lines indicate drift and diffusion of the reference process (where mini-batching is off) along PC1. 
    We observe that there are two quasi-equilibrium points, where $\langle\partial_t x_1\rangle\approx0$. 
    The stable fixed point is however at a much larger temperature than the marginally stable one: this temperature gradient is strong enough to trap the system at the marginally stable fixed point.}
\end{figure}

The demographic noise has an effect in this second stage of the dynamics, when the system is confined to the subset of high fitness states. If the population size is not too large, the demographic noise tends to skew the population composition towards any one of the fit genotypes. On these pure states, the demographic noise is indeed vanishing: this fact makes them thermophoretically favored, even though they are only nearly marginally stable from a deterministic standpoint. This classical result can be seen as a noise-induced transition \cite{transitions-book}, as reviewed in the SI \cite{SI}\footnote{We remark that the noise-induced transition refers here to a bifurcation for the maxima of the state distribution. This is not due to a drift induced by the Ito-Stratonovich conversion (as the SDE is already interpreted in the Ito sense) but to the boundedness of the domain.}.
Taking now individual fitness variability into account ---precisely, assuming that the level of variability of different species is not the same--- introduces an inhomogeneous rescaling of the demographic noise, which roughly tends to bias the steady-state distribution of the population towards states of lower average variance $\bar V(\boldsymbol z)$ (or higher average fitness $\bar f(\boldsymbol z)$, depending on their mutual tradeoff). 

Although this analysis applies to systems of any number of quasi-species $S$, the effect of annealed population heterogeneity is particularly pronounced in high-dimensional systems ($S\gg3$). As visible in Fig. \ref{fig:2} D, H and Figs. S6-S7, new peaks emerge in locations of the $(\bar f, \bar V)$ plane where the standard demographic noise from the reference process never concentrates the system so sharply. This enhancement of the non-equilibrium effect can be attributed to the role of entropic contributions in high dimensions. Since we are projecting from the higher dimensional space of population sates to the low-dimensional space of the summary statistics of interest, the changes we observe in the stationary distribution in the $(\bar f, \bar V)$ plane are no longer simply the manifestation of a noise-induced transition, but of a noise-induced \emph{phase} transition \cite{order-from-noise}.

In the example illustrated in Fig.~\ref{fig:3}, the steady-state dynamics appears quasi-1D, with two effective fixed points: an effective marginally stable one, and an effective stable one. Crucially, the amplitude of the fluctuations at the stable point substantially exceeds the intensity of the noise acting at the marginally stable point, a difference largely due to the mini-batching correction to the noise term. As shown in Fig. S8, the stable point is indeed corresponding to population states dominated by the genotype that is fittest in the average environment, which is also the one with the largest fitness variance (top corner of Fig. \ref{fig:2} C). In contrast, the marginally stable point corresponds to multiple types of microscopic population states, which are dominated by genotypes with lower mean fitness and lower variance (on the right side of the convex hull in Fig. \ref{fig:2} C). 
This explains the difference in local ``demographic temperature'' which is responsible for stabilizing the meta-stable state, effectively creating a non-equilibrium trap \cite{kantz-multiplicative}.

\section*{Discussion}
We have shown that when different individuals in a population repeatedly encounter distinct microenvironments, and these exposures reshuffle across generations, evolution is naturally biased toward genotypes that perform reliably across those varied challenges. This annealed heterogeneity produces an effect analogous to structured mini-batching in stochastic gradient descent: each generation supplies many small, domain-specific samples of the environment, and the resulting fluctuations act as an implicit form of regularization. In the language of learning theory, this amounts to a kind of across-domain generalization, favoring genotypes that capture features common to all microenvironments and that therefore remain fit even when facing novel environmental conditions not previously encountered. 

Through analytic work, we showed that this individual-level variability introduces a new source of demographic noise whose strength depends on the variance of fitness across microenvironments. In the diffusion limit, this noise effectively rescales the population size and penalizes genotypes whose fitness varies strongly from one microenvironment to another. Numerical experiments confirmed that, at intermediate population sizes, this bias produces macroscopic consequences: the population’s steady state shifts toward genotypes with lower across-environment variance and hence more uniform, generalist performance. 

%The Wright-Fisher model with individual mini-batching offers a tractable framework for examining how annealed population heterogeneity ---which can be realized, for example, via single-cell variability, or short-scale environmental heterogeneities--- influences population dynamics. Using the diffusion approximation, we analytically characterized the effective noise deriving from the stochastic protocol of individual mini-batching, and we provided a dynamical interpretation of how this acts as an implicit regularization mechanism, enhancing evolutionary robustness.

%Importantly, the effect of this implicit regularization turns out to be strong only at intermediate population sizes ---large enough for the diffusion approximation to be valid, but small enough to have a breakdown of the perturbative expansion of the nonlinear SDE. However, it is precisely at intermediate values of $N$ that having large variability represents a threat for a population, since it can more easily lead to extinction, and it is then most important for the system to be robust to stochastic fluctuations (i.e., suppress the variability). Our analysis therefore suggests that this form of non-inheritable variation in the fitness trait, described as individual mini-batching, could be seen as a self-regulating regularization mechanism, which turns off and on as the population size grows or decreases.

This phenomenon observed at intermediate population sizes can be understood as a noise-induced (phase) transition. Two key conditions are necessary for this transition to occur: multiple near-equilibria separated by low or no barriers must exist at the level of the average fitness landscape, and large ``temperature'' gradients must be realized within this network of near-equilibria. The occurrence of these two conditions clearly depends on the geometry of the genotypic space and the constitutive relation between the mean fitness landscape and the \emph{variance landscape} defined on this space. High-dimensional genotype-to-phenotype maps exhibiting soft modes or expressing low-dimensional constraints \cite{chris-review} are particularly likely to satisfy the first condition. Indications of these properties have been detected in numerous datasets, based on high-throughput experimental characterization of specific phenotypes \cite{otwinowski-global-ep, Kabir-global-ep,Desai-global-ep, Shenshen-global-ep,ArdLuis-neutral-pheno, Krug-landscapes}.
Yet, in order to predict whether competitive evolutionary dynamics can lead to the production of generalists in any given system, these measurements must be supplemented with a characterization of the fitness variability, which can be achieved through different approaches, depending on the source of noise that we aim to capture. For instance, a first approach could consist in building more detailed biophysical models of the processes generating the discussed heterogeneities ---e.g. modeling the individual reactions of complex multi-step selection processes occurring in a highly heterogeneous environment, like affinity maturation, or constructing relatively low-dimensional, noisy genotype-to-phenotype-to-fitness maps by marginalizing over larger genetic interaction networks \cite{omnigenic-model,omnigenic-inheritance,Petrov-fitness-modularity}. Variance in the fitness distributions can indeed be originating not only from exterior heterogeneity, but also from internal variability which makes different individuals perceive the same environment differently. Alternatively, this characterization could be achieved through direct experimental measurements ---e.g. via experimental assays measuring overdispersion of division times \cite{mother-machine}, extrinsic expression noise \cite{Elowitz-noise-science, Elowitz-Siggia-extrinsic, Shalek-microfluidic}, or demographic noise in single-species colonies \cite{Hallatschek-Nelson-KPZ,Hallatschek-demographic}.

Interestingly, significant differences in demographic noise have been observed among \emph{E. coli} colonies that differ by single gene deletion mutations, even when grown in identical conditions \cite{Hallatschek-demographic}, or among colonies of the same strain grown at different temperatures \cite{Hallatschek-surfing}. Precisely, it was found in \cite{Hallatschek-demographic} that these demographic noise differences correlate with colony-level traits related to overall growth rate, but not with cell-level traits ---with faster growing colonies exhibiting larger demographic noise. %Notably, it was observed that faster-growing colonies exhibit larger demographic noise. 
Our model suggests that these differences in the strength of demographic noise across colonies could originate from a different level of phenotypic variability in each strain. 
Since in this experiment each colony is made of an isogenic population, the amplitude of the demographic noise in each colony must indeed be a constant and reflect the mean-variance relation in the fitness distribution of the strain. The observation then suggests that the rescaling factor $V/f^2$ associated to each strain should be monotonically related to the overall growth rate of the colony, $f$. A simple mechanistic hypothesis justifying this relation is that rapidly dividing cells might dilute their protein content more extensively, causing greater fluctuations in relative protein concentrations, which may lead to increased variance in fitness-related traits. %our theoretical framework could be extended to make more precise predictions about demographic noise in freely expanding populations where population size is not held constant. Such adaptations would provide better alignment between theoretical predictions and experimental observations in natural growth conditions.
This conjecture ---apparently at odds with previous findings relating growth rate to expression noise and phenotypic variability \cite{stoch-metabolism-nature,vanNimwegen-PlosBio,vanNimwegen-pnas-effective-bet-hedging, SprattLane-review, SprattLane-exp}--- opens avenues for future investigations: for example, it would be valuable to experimentally verify the extent to which colony-level demographic noise measurements can predict extrinsic noise levels in fitness-related genes, by systematically examining how gene knockouts affect transcriptional noise levels. 
Our theoretical framework, by connecting demographic noise levels with variability in growth rates, could help interpret and design experiments to better understand how noise is propagated and fed back in genotype-to-phenotype-to-fitness maps.

%Finally, it would be interesting to investigate how additional knobs, such as correlations in the population heterogeneity or slower annealing, influence the implicit regularization mechanism in Wright-Fisher and other population dynamics models, bridging the scales between intrinsic cell-to-cell variability and coherent environmental fluctuations, in order to provide a unifying view of how fluctuations shape selection.

In summary, our results illustrate how the texture of environmental experience, and not just its long-term average, can shape evolutionary outcomes. By controlling how different microenvironments are sampled across a population, we can bias evolution toward solutions that anticipate novelty without any explicit foresight. It is worth noting that the mathematical results obtained from the diffusion approximation apply beyond the scenario where generalists and specialists compete with each other: these include the case of isogenic populations described above, as well as the case where only specialists compete with each other (i.e. when there is no appreciable difference in the fitness variance of distinct genotypes). In the latter scenario, $V/\bar f(\boldsymbol z)^2$ will be shaped by the fitness dependence, and our analytical prediction is that the inductive bias will strengthen and accelerate the selection of the fittest genotypes in the average environment. It would be interesting to investigate how additional knobs, such as correlations in the population heterogeneity or slower annealing, influence the implicit regularization mechanism, bridging the scales between intrinsic cell-to-cell variability and coherent environmental fluctuations, in order to provide a unifying view of how fluctuations shape selection.

\section*{Materials and Methods}
For the numerical analysis presented in the main text of this paper we consider a training set of $M=100$ microenvironments corresponding to random realizations of the "interface-like" fitness model described in \ref{fig:2} A. Of the 16 possible distinct landscapes that can be realized for an environment vector of length $L=8$ with $K=4$ binary variable features, only 15 were realized, with non-uniform frequency.  
We simulate the evolutionary process with annealed population heterogeneity using the algorithm outlined in Fig.~\ref{fig:1} C; a detailed description can be found in the SI \cite{SI}. The steady state distributions and steady-state averages are obtained from the integration of 700 trajectories that start from 7 different types of initial conditions (isogenic populations localized on 5 random genotypes and on the $(0,\dots,0)$ and $(1,\dots,1)$ genotypes). The mutation rate adopted in the illustrated numerical example is of 0.005 mutations/site/generation.

For the three-species case presented in the SI, we generate $M=500$ samples of fitness values per genotype, drawn from log-normal distributions with mean $\mu_a$ and variance $\sigma_a^2$, obtaining the empirical mean $f_a$ and empirical variance $V_a$ reported in Fig. S1. The evolutionary algorithm is the same as in the interface-like model, with a uniform mutation rate of 0.001 between any pair of genotypes. Steady state observables are obtained from 50 replicates of the stochastic trajectories, where the initial condition is the balanced population state $z_a=z_b=z_c=1/3$.

\section*{Data, Materials, and Software Availability}
The code to reproduce all simulations and analysis has been deposited in Github (\url{https://github.com/baby-ff/mini-batching-evolution}). Additional datasets are available from the authors upon request.

\begin{acknowledgments}
We acknowledge useful conversations with Michel Fruchart, Omer Granek, and members of the Chakraborty, Murugan and Ranganathan groups.
F.F. was supported by the NSF-Simons National Institute for Theory and Mathematics in Biology (NITMB) Fellowship supported via grants from the NSF (DMS-2235451) and Simons Foundation (MPS-NITMB-00005320).
\end{acknowledgments}

\appendix
\renewcommand\thefigure{S\arabic{figure}}    
\setcounter{figure}{0}    

\section{Diffusion approximation}
In this section we provide a detailed derivation of the SDE describing the evolutionary process with individual variability in the diffusion approximation, Eqs. 2--4 in the main text.
In order to model the effect of extrinsic noise in the growth rates of the populations onto the evolutionary dynamics, we start by considering a composite Markov chain where, at each generation, the system undergoes the following sequence of steps:
\begin{enumerate}
    \item \textbf{Mini-Batching.} Suppose that, at time $t$, the population is composed of $n_a$ individuals for each quasi-species $a=1,\dots, S$. We identify here the quasi-species with classes of individuals sharing the same genotypes, which we assume to be noiselessly inherited in the absence of mutations. However, the genotype-to-fitness map is assumed to be stochastic, and i.i.d. across isogenic individuals, i.e. $\forall i=1,\dots,n_a$, $f_i \sim P_a(f)$. We indicate with $f_a$ and $V_a$ the mean and the variance of the distribution, respectively. The ``mini-batching'' step consists in assigning a fitness value to each individual in the population. For the sake of simplicity, we consider a discrete set of fitness values $\{f_a^\mu\}_{\mu=1,\dots, M}$, chosen with probabilities $\{\mathcal P_a^\mu\}$, but the same derivation can be obtained for fitness values defined on continuous domains.
    Therefore, the batching step is described as $S$ independent multinomial sampling processes (one per genotype class), whose result is a set $\{n_a^\mu\}_{a=1\dots S;\mu=1\dots M}\in \mathcal S$:
    \begin{equation}
        B\left(\{n_a^\mu\}_{a=1,\dots, S;\mu=1,\dots, M}\vert \{n_a\}\right) = \prod_{a=1}^S B_a\left(\{n_a^\mu\}_{\mu=1,\dots, M}\vert n_a\right) = \prod_{a=1}^S n_a!\prod_{\mu=1}^M\frac{\left(\mathcal P_a^\mu\right)^{n_a^\mu}}{n_a^\mu!},
    \end{equation}
    where $\mathcal S$ is the simplicial complex with $S-1$ faces $\Delta_{M-1}$, obtained from the constraints $\sum_\mu n_a^\mu=n_a$ and $\sum_a n_a = N$, In this model, $N$ is the fixed total population size.
    \item \textbf{Wright-Fisher selection.} Once the fitness of each individual in the population is specified, we describe their all-to-all competition via the Wright-Fisher (WF) dynamics:
    \begin{equation}
        W\left(\{n_a^\mu\}'_{a;\mu}\vert\{n_a^\mu\}_{a;\mu}\right) = N!\prod_{a=1}^S \prod_{\mu=1}^M\frac{(\pi_a^\mu)^{(n_a^\mu)'}}{(n_a^\mu)'!},
    \end{equation}
    where the multinomial probability parameter corresponds to the relative fitness, $\pi_a^\mu = \frac{f_a^\mu n_a^\mu}{\sum_{b\nu}f_b^\nu n_b^\nu}$.
    \item \textbf{Resetting.} Since, in our description, all the perfectly inherited traits are incorporated in the quasi-species label $a$, we want to assign, after the replication, a new independent fitness value to each individual (given its genotype). Therefore we \emph{erase }the $\mu$ labels by regrouping the subpopulations into a vector $\{n_a\}'$ s.t. $n_a' = \sum_\mu (n_a^\mu)'$ $\forall a$. Hence the propagator associated to this resetting step reads:
    \begin{equation}
        R\left(\{n_a\}'\vert \{n_a^\mu\}'\right) = \prod_{a=1}^S \delta\left(n_a' - \sum_\mu(n_a^\mu)'\right).
    \end{equation}
    \item \textbf{Genetic mutations.} We finally allow for genetic mutations to occur, where the probability to mutate from a genotype $a$ to a genotype $b$ per generation is expressed as $\nu\Lambda_{ba}$, where $\nu$ is a scale parameter and $\Lambda$ is the weighted Laplacian of the mutational graph. We indcate the transition probability associated to this last step as 
    \begin{equation}
        G_\Lambda\left(\{n_a\}''\vert \{n_a\}'\right).
    \end{equation}
\end{enumerate}
The resulting transition probability between two subsequent generations can then be formally computed as a convolution of the transition probabilities for the Markov sub-steps 1-4:
\begin{equation}
    P\left(\{n_a\}''\vert \{n_a\}\right) = G_\Lambda \left(\{n_a\}''\vert \{n_a\}'\right) * R\left(\{n_a\}'\vert \{n_a^\mu\}'\right) * W\left(\{n_a^\mu\}'_{a;\mu}\vert\{n_a^\mu\}_{a;\mu}\right) * B\left(\{n_a^\mu\}_{a;\mu}\vert \{n_a\}\right).
    \label{composite-Ptr}
\end{equation}

We now use \eqref{composite-Ptr} to derive the stochastic dynamic equations of the process in the diffusion approximation. We recall that, given a sequence of Markov chains $\{X_t^N\}_{t\geq0}$, with $\delta X_t^N = X_{t+1}^N - X_t^N$, the diffusion approximation holds, in the $N\to\infty$ limit, when the following conditions are satisfied:
\begin{enumerate}
    \item $\mathbb E\left[\delta X^N_t\vert X^N_t\right] = h_N D_1 (X_t^N) + \epsilon_{1,t}^N,$
    \item $\mathbb E\left[\left(\delta X^N_t\right)^2\vert X^N_t\right] = h_N D_2 (X_t^N) + \epsilon_{2,t}^N,$
    \item $\mathbb E\left[\left(\delta X^N_t \right)^k\vert X^N_t\right] = \epsilon_{k,t}^N \quad \mathrm{for}\ k>2,$
\end{enumerate}
where $h_N\to 0^+$ as $N\to\infty$ and $\forall \tau>0$, $\sum_{t<\lfloor {\tau/h_N}\rfloor}\lvert\epsilon_{k,t}^N\rvert$ $\forall k$. Then, the discrete process $\{X_{t=\lfloor\tau/h_N\rfloor}^N\}_{t\geq 0}$ converges in distribution to a continuous-time diffusion process $\{X(\tau)\}_{t\in\mathbb R^+}$ with drift $D_1(X_t)$ and variance $D_2(X_t)$.
In other words, the diffusion approximation holds when the distribution of the chain increments can be approximated with a Gaussian.
In the case of WF processes, where $N$ represents the population size, the diffusion approximation is known to be valid when the fitness advantage of each quasi-species present in the population is infinitesimal (as well as the mutation rates, if mutations are incorporated in the model). 

We proceed to compute the first moments of the chain increments using the composite transition probability \eqref{composite-Ptr}. For simplicity, we ignore in this derivation the genetic mutations introduced in the last step, since ---as we will show at the end--- these do not affect our main result, i.e. the characterization of the effective noise due to the annealed population heterogeneity.

\emph{First moments}
\begin{equation}
\mathbb E_{R*W*B}\left[z_a' - z_a\vert \{z_a\}\right] = \frac{1}{N}\mathbb E_B\left[\sum_\mu \mathbb E_W\left[(n_a^\mu)'\vert \{n_a^\mu\}\right]\big\vert \{n_a\}\right]-z_a = \mathbb E_B\left[\sum_\mu\pi_a^\mu\big\vert\{n_a\}\right]-z_a.
\end{equation}
Define 
\begin{equation}
    \delta n_a^\mu = n_a^\mu - \mathbb E_B\left[n_a^\mu\vert\{n_a\}\right] = n_a^\mu - \mathcal P_a^\mu n_a ,
\end{equation}
such that (shortening the notation for the conditional)
\begin{equation}
    \mathbb E_B\left[\delta n_a^\mu\right] = 0;\qquad \mathbb E_B\left[\delta n_a^\mu\delta n_b^\nu\right] = \delta_{ab}n_a\left( \delta_{\mu\nu}\mathcal P_a^\mu - \mathcal P_a^\mu\mathcal P_a^\nu\right).
\end{equation}
Rename 
\begin{equation}
    \pi_a^\mu = \frac{}{} = \frac{\mathcal N_a^\mu + \delta \mathcal N_a^\mu}{\mathcal D+ \delta \mathcal D} = \frac{\mathcal N_a^\mu}{\mathcal D} +  \frac{\delta \mathcal N_a^\mu}{\mathcal D} - \frac{\delta \mathcal N_a^\mu\delta \mathcal D}{\mathcal D^2} +  \frac{\mathcal N_a^\mu\delta \mathcal D^2}{\mathcal D^3} + \dots
\end{equation}
where 
\begin{equation}
    \mathcal N_a^\mu = f_an_a\mathcal P_a^\mu, \qquad \mathcal D = \sum_b f_b n_b = N \bar f(z), \qquad \delta \mathcal N_a^\mu = f_a^\mu\delta n_a^\mu,\qquad \delta \mathcal D = \sum_{b,\nu} f_b^\nu \delta n_b^\nu
\end{equation}
such that
\begin{eqnarray}
    &\mathbb E_B\left[\delta \mathcal N_a^\mu\right]=\mathbb E_B\left[\delta \mathcal D\right]=0,\qquad &\mathbb E_B\left[\delta \mathcal N_a^\mu\delta \mathcal N_b^\nu\right]=\delta_{ab} n_af_a^\mu f_a^\nu\left(\delta_{\mu\nu}\mathcal P_a^\mu - \mathcal P_a^\mu\mathcal P_a^\nu\right),\\ 
    &\mathbb E_B\left[\delta\mathcal N_a^\mu \delta \mathcal D\right]=(f_a^\mu)^2n_a \mathcal P_a^\mu (1-\mathcal P_a^\mu), \qquad &\mathbb E_B\left[\delta \mathcal D^2\right]=\sum_a V_a n_a = N\bar V(z), 
\end{eqnarray}
and the neglected terms $(\dots)$ are of higher order in $N^{-1}$. Therefore 
\begin{equation}
    \mathbb E_{R*W*B}\left[z_a' - z_a\vert \{z_a\}\right] = \frac{f_a - \bar f(\boldsymbol z)}{\bar f(\boldsymbol z)}z_a - \frac{1}{N}\frac{V_a- \bar V(\boldsymbol z)}{{\bar f(\boldsymbol z)}^2}z_a + o(N^{-1}).
\end{equation}
The diffusion approximation condition (i) is satisfied if $\frac{f_a - \bar f(z)}{\bar f(z)}z_a\to 0^+$ as $N\to\infty$. While this condition might not be valid in the initial transient of the population dynamics, it will be asymptotically valid as the system approaches one of the pure states corresponding to the vertices of the $\Delta_{S-1}$ simplex. Let us identify $h_N=1/N$ and define the drift term of the Langevin process:
\begin{equation}
    F_a(\boldsymbol z) = \lim_{N\to\infty}\frac{\mathbb E_{R*W*B}\left[z_a' - z_a\vert \{z_a\}\right]}{1/N}=\underbrace{N\frac{f_a - \bar f(\boldsymbol z)}{\bar f(\boldsymbol z)}z_a}_{F^0_a(\boldsymbol z)}-\frac{V_a-\bar V(\boldsymbol z)}{\bar f(\boldsymbol z)^2}z_a,
\end{equation}
where $F^0_a(\boldsymbol z)$ is the drift of the reference process describing the evolution of the system in the average environment, where each quasi-species $a$ is assigned a unique fitness value $f_a$.

\emph{Second moments}
\begin{multline}
    \mathbb E_{R*W*B}\left[\left(z_a' - z_a - F_a(\boldsymbol z)/N\right)\left(z_b' - z_b - F_b(\boldsymbol z)/N\right)\vert \{z_a\}\right] = \\
    \frac{1}{N^2}\mathbb E_B\left[\sum_{\mu,\nu} \mathbb E_W\left[\left((n_a^\mu)' - \pi_a^\mu N\right)\left((n_b^\nu)' - \pi_b^\nu N\right)\vert \{n_a^\mu\}\right] + \left(\pi_a^\mu N - n_a-NF_a\right)\left(\pi_b^\nu N - n_b - N F_b\right)\right] =\\
    \frac{1}{N^2}\mathbb E_B\left[N\left(\delta_{ab}\sum_\mu\pi_a^\mu-\sum_{\mu\nu}\pi_a^\mu\pi_b^\nu\right) + \left(\pi_a^\mu N - n_a-NF_a\right)\left(\pi_b^\nu N - n_b - N F_b\right)\right] = \\
    \hat \pi_a - \hat\pi_a\hat\pi_b + O(N^{-1}) + \frac{1}{N^2}\mathbb E_B\left[\left(\pi_a^\mu N - n_a-NF_a\right)\left(\pi_b^\nu N - n_b - N F_b\right)\right] =\\
    \hat \pi_a - \hat\pi_a\hat\pi_b + \delta_{ab}\frac{V_az_a}{\bar f^2} + \hat\pi_a \frac{f_b \bar V - V_b \bar f}{\bar f^3}z_b + \hat \pi_b\frac{f_a\bar V - V_a\bar f}{\bar f^3}z_a + O(N^{-1})
,
\end{multline}
where $\hat \pi_a = \sum_\mu\mathbb E_B[\pi_a^\mu] = F_a/N + z_a = f_az_a/\bar f$. Analogously to the drift, we can compute the diffusion tensor of the process
\begin{multline}
    D_{ab}(\boldsymbol z) = \lim_{N\to\infty}\frac{\mathbb E_{R*W*B}\left[\left(z_a' - z_a - F_a(\boldsymbol z)/N\right)\left(z_b' - z_b - F_b(\boldsymbol z)/N\right)\vert \{z_a\}\right]}{1/N} = \\\underbrace{\left(\delta_{ab}z_a - z_a z_b\right)}_{D^0_{ab}(\boldsymbol z)}\left(1 + \frac{\bar V(\boldsymbol z)}{\bar f(\boldsymbol z)^2}\right) + \delta_{ab}z_a\frac{V_a- \bar V(\boldsymbol z)}{\bar f(\boldsymbol z)^2} - z_a z_b \frac{V_a + V_b - 2\bar V(\boldsymbol z)}{\bar f(\boldsymbol z)^2}
\end{multline}
where we have exploited the previous assumption (from condition (i)) that $\left[f_a/ \bar f(\boldsymbol z)-1\right]z_a\lesssim O(1/N)$. Here $D^0_{ab}(\boldsymbol z)$ is the diffusion tensor associated to the demographic noise of the reference process. 

It can be shown that the higher cumulants of the increments are negligible as $N\to\infty$, which satisfies condition (iii).

We can now reintroduce mutations (step 4) and consider the increments' distribution associated to the full composite Markov process. Explicitly writing the propagator $G_\Lambda$ is in general complicated, as it requires enumerating all the possible ways to go from an initial state $\{n_a'\}$ to a final state $\{n_a''\}$ via combinations of mutations. We therefore restrict to a weak mutation limit where the probability of having more than one mutation per individual per generation can be neglected. Let us denote $\Lambda_{ab}$ the graph laplacian of the graph connecting all pairs of sequences at a mutational distance of 1, and let $\nu$ indicate the (small) mutation rate. In this regime, 
$\mathbb E_{G_\Lambda}[n_a'' - n_a'|\{n_a'\}]=\nu \Lambda_{ab}n_b'$, so
\begin{multline}
    \mathbb E_{G_\Lambda*R*W*B}[z_a'' - z_a|\{z_a\}] = \frac{1}{N} \mathbb E_{G_\Lambda*R*W*B}[n_a'' - n_a'|\{n_a\}]+\frac{F_a(\boldsymbol z)}{N} 
    =\frac{1}{N} \mathbb E_{R*W*B}\left[\nu\Lambda_{ab} n_b'\vert \{n_a\}\right]+\frac{F_a(\boldsymbol z)}{N} \\=  \nu\Lambda_{ab}\frac{f_b}{\bar f(\boldsymbol z)} z_b +O(\nu N^{-1})+\frac{F_a(\boldsymbol z)}{N}=\frac{\Delta F_a(\boldsymbol z)}{N} + \frac{F_a(\boldsymbol z)}{N} + O(\nu N^{-1}). 
\end{multline}
For the diffusion approximation to be valid, in addition to the previous conditions, we must require that $\nu \lesssim O(N^{-1})$. Exploiting this fact, we can neglect the last term in the previous equation and show that there are no corrections to the second cumulants of the increments:
\begin{multline}
    \mathbb E_{G_\Lambda*R*W*B}\left[\left(z_a'' - z_a - F_a(\boldsymbol z) - \Delta F_a(\boldsymbol z)\right)\left(z_b'' - z_b - F_b(\boldsymbol z) - \Delta F_b(\boldsymbol z)\right)|\{z_a\}\right] = \\
    \mathbb E_{R*W*B}\left[\mathbb E_{G_\Lambda}\left[\left(n_a'' - n_a' - \Delta F_a(\boldsymbol z)/N\right)\left(n_b'' - n_b' -  \Delta F_b(\boldsymbol z)\right)/N|\{n_a'\}\right]\big\vert\{n_a\}\right]+\frac{D_{ab}(\boldsymbol z)}{N} =  \frac{D_{ab}(\boldsymbol z)}{N}  + O(\nu N^{-1}).
\end{multline}
Finally, the SDE describing the evolutionary dynamics with mini-batching reads
\begin{equation}
    \partial_t z_a = F_a(\boldsymbol z) + B_{ab}(\boldsymbol z) \xi_b(t), \qquad F_a(\boldsymbol z) = F^0_a(\boldsymbol z) + \Delta F_a(\boldsymbol z)  ,\quad \left[B(\boldsymbol z)B^\top(\boldsymbol z)\right]_{ab} = D_{ab}(\boldsymbol z).
\end{equation}
The SDE has to be integrated with Ito's prescription.
The correctness of the formulas obtained has been checked in the simple $S=2$ case, where the process is one-dimensional.

This Langevin equation has been used to derive the variants of the Price equation in Eqs. --\eqref{Price-V}, using the definition of the population averages of mean and variance of the fitness distributions: $\bar f = \boldsymbol f^\top \boldsymbol z$, $\bar V = \boldsymbol V^\top \boldsymbol z$. We report here the full analytical expressions of the diffusion matrix, omitted from the main text for clarity:
\begin{eqnarray}
    \langle\eta_f\eta_f\rangle & =\boldsymbol f^\top D \boldsymbol f &= Var_z(f)\left(1 + \frac{\bar V}{\bar f^2}\right)+ \frac{Cov_z\left((f-\bar f)^2,V-\bar V\right)}{\bar f^2},\\  
    \langle \eta_V\eta_V\rangle & =\boldsymbol V^\top D \boldsymbol V &= Var_z(f)\left(1 + \frac{\bar V}{\bar f^2}\right)+ \frac{\overline{(V-\bar V)^3}}{\bar f^2},\\  
    \langle \eta_f\eta_V \rangle & =\boldsymbol f^\top D \boldsymbol V &= Cov_z(f,V)\left(1 - \frac{\bar V}{\bar f^2}\right) + \frac{Cov_z(f,V^2)}{\bar f^2}.
\end{eqnarray}
The generalized Price equation for any collective observable $\bar u(\boldsymbol{z}) = \sum_az_au_a$ (e.g. the Price equation describing the evolution of the system along the first principal component, illustrated in Fig. 3 C) can be obtained as
\begin{equation}
    \partial_t \bar u(\boldsymbol{z})= N\frac{ Cov_z(u,f) + \nu \boldsymbol f^\top \Lambda \tilde {\boldsymbol{z}}}{\bar f} - \frac{Cov_z(u,V)}{\bar f^2}+\eta_U
    \label{Price-A}
\end{equation}
where 
\begin{equation}
    \langle\eta_U \eta_U\rangle = \boldsymbol u^\top D\boldsymbol u = T_U
\end{equation}

\subsection*{The three-species case}

For a state space with an arbitrary number of species, the generalized Price equations Eq. 6 and Eq. 7 do not form a closed system of equations, but for $S=3$ this becomes a closed two-dimensional system, thanks to an invertible mapping from the population state $\boldsymbol z$ to the collective observables $\bar f$ and $\bar V$ ---unless the three points representing the three species in the $(f,V)$ plane are collinear. 

In this section we analyze the dynamics of the system in this simple low-dimensional scenario, where things are easy to visualize. Consider an illustrative example where $f_a\simeq f_b>f_c$ and $\nu\ll 1$, corresponding to an evolutionary regime where mutations are rare and there exists a (nearly) neutral network of high fitness genotypes ($a$ and $b$).
As shown in Fig.~\ref{fig:3spec} A, there is typically a separation of scales in the deterministic dynamics: first, the system experiences the fast rightward push dictated by the leading order term in Eq. 6, followed by the slow relaxation along the vertical axis towards the mutation-selection-balance fixed point. The demographic noise mostly acts in this second stage of the dynamics, by skewing the population composition towards either one of the two fit genotypes, $a$ or $b$.
Fig. \ref{fig:3spec} B shows how the introduction of mini-batching clearly moves the noise-induced transition point to larger population sizes and enhances the occupation of the low-variance states; however, in the considered $S=3$ example, it is hard to disentangle the effect of the noise asymmetry from that of the deterministic correction. A more striking effect is shown in the high-diemnsional case discussed in the main text and reported in Figs.~\ref{fig:S10} and \ref{fig:S9}.

\begin{figure}
    \centering
    \includegraphics[width=0.95\linewidth]{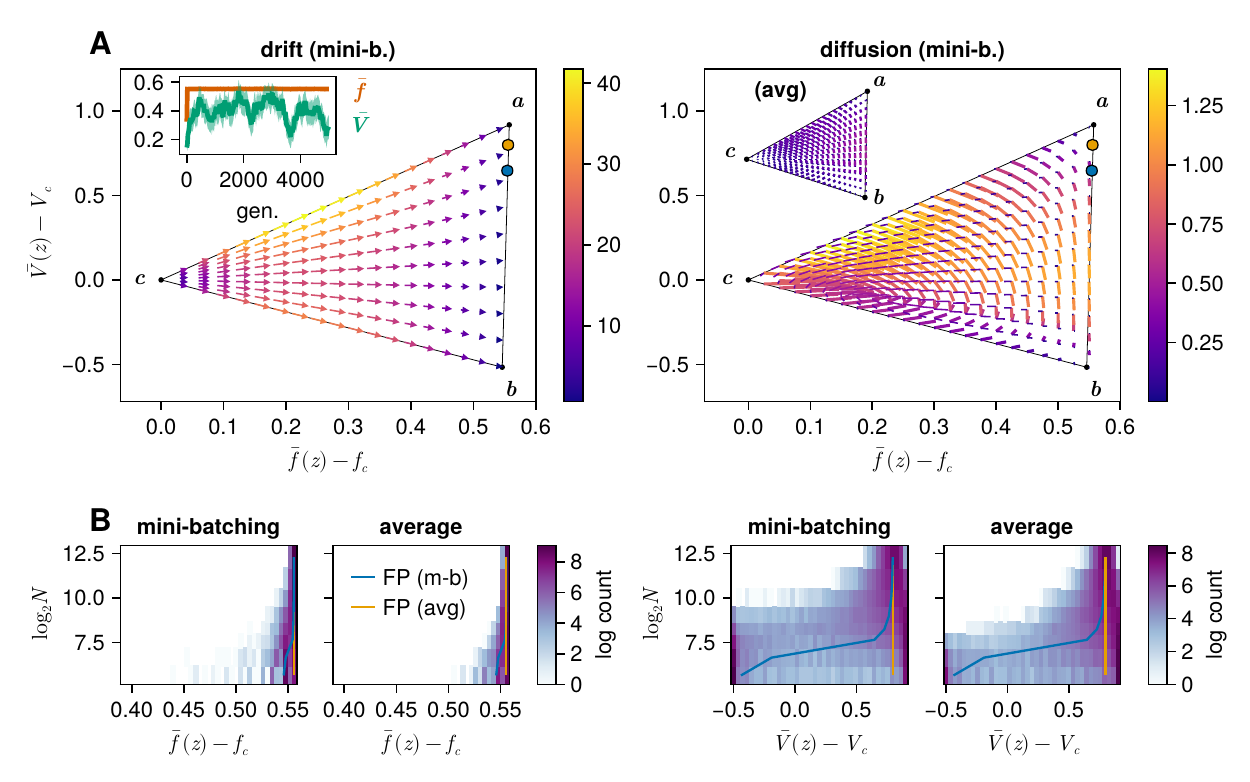}
    \caption[S1]{\textbf{Three-species model shows noise-induced transitions.} \textbf{A Drift and diffusion terms of the Ito-SDE.} Drift vector field (\emph{left}) and diffusion tensor glyph (\emph{right}) representing the strength of the deterministic and stochastic forces acting on the three-species system. The colormap shows a band of large diffusivity in the whole region where the $a$ and $b$ species coexist, elucidating the mechanism by which the system is pushed towards the two rightmost vertices of the simplex. The $N$-dependent drift is here computed for $N=200$. Blue and yellow points represent the fixed points of the deterministic counterparts of the SDEs Eq. 3--Eq. 4. \emph{Left inset}: Time evolution of the collective observables $\bar f(\boldsymbol z)-f_c$ and $\bar V(\boldsymbol z)-V_c$, averaged over 50 trajectories, with initial conditions $z_a\approx z_b \approx z_c \approx 1/3.$ The fitness axis identifies the fast manifold of the dynamical system, while the variance axis coincides with the slow manifold, along which the system wanders in the presence of noise. \emph{Right inset}: Tensor glyph for the diffusion of the reference process. The direction of the eigenvectors (represented by the rods) does not seem change much with respect to the main plot, while the amplitude of the associated eigenvalues does (denoted by rod length and color). This is in agreement with the assumption that the correction to the diffusion term is mostly captured by an isotropic rescaling of the tensor in Eq. 4.  \textbf{B Noise-induced transition.} As in Fig. 2, the steady-state distributions of the population mean fitness and of the population variance are affected by the population size: the number of modes decreases from two to one as $N$ increases. This transition happens both in the reference 
    }
    \label{fig:3spec}
\end{figure}

\begin{figure}[h]
    \ContinuedFloat
    \caption{case and in the presence of individual mini-batching, with the bimodality retained up to larger values of $N$ in the presence of mini-batching.
    We remark that the observed transition is not associated to a bifurcation in the deterministic counterpart of the model: the number of fixed points (FP - solid lines) remains one for any value of $N$. 
         }
\end{figure}

\section{Noise-induced transition in nearly-neutral theory of evolution}
For the sake of completeness, we review the basic mechanism by which demographic noise produces a noise-induced transition in the simple setting where two species compete with each other in evolution. Let us assume that one of the species has a small fitness advantage over the other, that the mutation rate is small, and that the genotype-to-fitness map is not stochastic. Under these assumptions, the dynamics of the system in the diffusion approximation is described by the following Ito-SDE:
\begin{equation}
    \partial_t y = N\left\{ \frac{s}{1 + s y}\left(\frac{1}{2}-y\right)\left(\frac{1}{2}+y\right) - \frac{2\nu y}{1+sy}\right\} + \sqrt{\left(\frac{1}{2}-y\right)\left(\frac{1}{2}+y\right)}\xi,
    \label{y-SDE}
\end{equation}
where $y = z-1/2$, $z$ is the fraction of individuals of the first species in the population, $s=2(f_a-f_b)/(f_a+f_b)$ is the relative fitness advantage of the first species with respect to the second, and $\nu$ is the symmetric mutation rate between the two species. Notice that $y\in[-1/2,1/2]$, with natural boundary conditions. The fixed points of the system will be the solutions of the quadratic equation coming from the drift term that live in this interval. It can be shown that for $\nu>0$ there always exists a single solution of this kind, corresponding to 
\begin{equation}
    y^* = \begin{cases}
    0\qquad\qquad\qquad\qquad \text{if } s=0\\
    -\frac{\nu}{s} + \sqrt{\frac{\nu^2}{s^2} + \frac{1}{4}}\quad\  \text{if } s\neq0
    \end{cases}
    \label{y*}
\end{equation}
When $\nu=0$, two solutions exist: $y^*=\pm 1/2$ .
We want to determine the local maximum point(s) $\tilde y$ of the steady-state probability distribution $P(y)$ and compare them to $y^*$. 

The transition occurs when $\tilde y$ differs from $y^*$, and especially when the number of extrema of the p.d.f. differs form the number of stable fixed points of the deterministic system. Let us notice that when the noise is additive and the system has natural boundary conditions, the noise cannot move the position of the maximum of the probability distribution away from the fixed points of the deterministic dynamics. In general, when the noise is multiplicative, the extrema of the steady-state probability distribution do not coincide with the dynamical fixed points, but the conditions for a bifurcation are difficult to study in arbitrary systems. For the system in \eqref{y-SDE}, since the process is one-dimensional, the solution of steady-state Fokker Planck equation (FPE) is easy to find. The FPE can be written as a continuity equation, $\partial_y J(y) = 0$, whose solution in 1D corresponds to $J(y) = J$ (constant). The probability current reads 
\begin{equation}
    J(y)=N\left\{ \frac{s}{1 + s y}\left(\frac{1}{4}-y^2\right) - \frac{2 \nu y}{1+sy}\right\}P(y) -\frac{1}{2} \partial_y \left\{\left(\frac{1}{4}-y^2\right) P(y)\right\}. 
\end{equation}
The natural boundary conditions $J(\pm 1/2)=0$ impose $J=0$, and the solution of the Ito-FPE reads
\begin{equation}
    P(y)\propto \left(1+sy\right)^{2N\left(1-\frac{8\nu}{4-s^2}\right)}\left(1-2y\right)^{\frac{4N\nu}{2+s}-1} \left(1+2y\right)^{\frac{4N\nu}{2-s}-1} .
    \label{Py}
\end{equation}
Notice that, despite $1\pm 2y\to 0$ for $y\to\pm1/2$, the previous expression is always normalizable for $|s|<1$ (in fact $s\ll1$ for the diffusion approximation to be valid). When $N<N_-=\frac{2-|s|}{4\nu}$, the extrema of $P(y)$ are clearly at the boundaries of the domain, $y=\pm1/2$, where the function diverges. When $N_-<N<N_+=\frac{2+|s|}{4\nu}$, the function only diverges on one extremum of the domain ($y=1/2$ if $s>0$, or viceversa). 
In contrast, when $N>N_+$ $\lim_{y\to\pm1/2}P(y)=0$, so we must look for the extrema of the p.d.f. in the interior of the domain. Hence $\tilde y$ will satisfy $\partial_y \ln P(\tilde y)=0$, i.e.,
\begin{equation}
    Ns \left(1-\frac{8\nu}{4-s^2}\right)\left(1-4\tilde y^2\right) + 4\tilde y (1+s\tilde y)+\frac{8N \nu}{4-s^2}(1+s\tilde y)(s-4\tilde y)=0.
    \label{tilde-y}
\end{equation}
There exists only one solution for \eqref{tilde-y} in the domain, if we further require that $\nu\leq 1/2$ (for the diffusion approximation to be valid):
\begin{equation}
    \tilde y = \frac{1-2N\nu+\sqrt{1+(N-1)Ns^2+4N\nu(N\nu-1)}}{2(N-1)s}
    \label{yext}
\end{equation}
which can be checked to be a maximum of $P(y)$ and to coincide with $y^*$ only when $N\to\infty$. 

\begin{figure}
    \centering
    \includegraphics[width=0.95\linewidth]{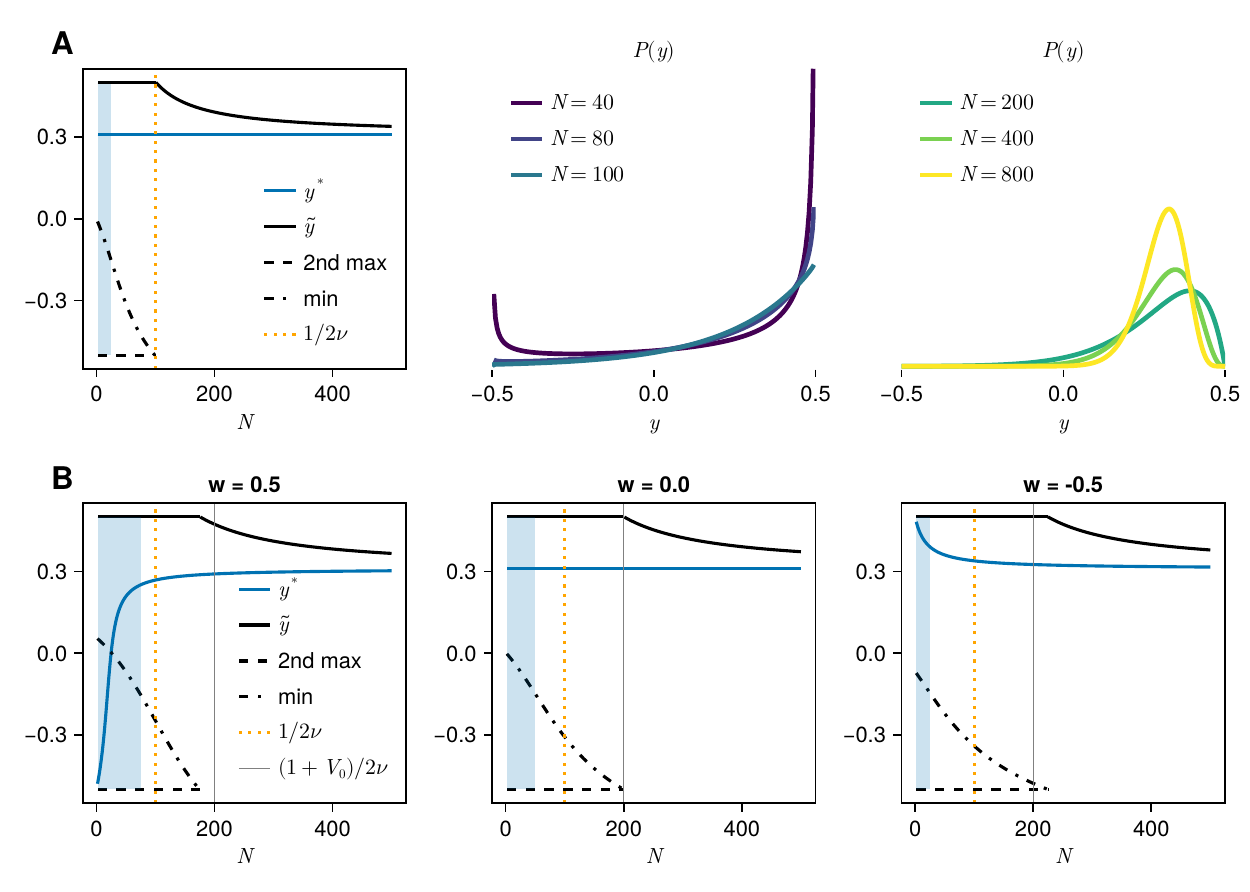}
    \caption{\textbf{Noise-induced phase transition (two-species system).} \textbf{A} Bifurcation vs phase diagram in the absence of mini-batching (average landscape), and associated probability densities for representative population sizes (cfr. \eqref{y*},\eqref{yext}, \eqref{Py}). \textbf{B} Bifurcation vs phase diagram in the presence of mini-batching. As highlighted in \eqref{fp-mb} and \eqref{ext-mb}, the fixed point is now also function of $N$ and $w$, and the transition point strongly depends on $V_0$, in addition to $w$, $s$ and $\nu$. The parameters used for this illustration are: $V_0=1$, $s=0.02$, $\nu=0.005$. The shaded area represents the region $N<N_c$, \eqref{eq:Nc}.}
    \label{fig:bifurc-diags}
\end{figure}

This recapitulates the noise-induced transition from the U-shaped distribution (for $N<N_c$) to a unimodal distribution centered around $\tilde y_N\to y^*$, as illustrated in Fig~\ref{fig:bifurc-diags}. Let us remark that the threshold population size at which the transition occurs only depends on the mutation rate between the two species, in the absence of single-cell variability. This threshold marks the transition from the regime where the deterministic mutation-selection balance condition is typically realized, to the case where it is only realized on average, while the typical population configurations are localized on any of the two species. This regime is the one described by the equilibrium-like theory reviewed in the next SI section. 

For a two-species system with single-cell variability, the solution of the steady-state FPE with natural boundary conditions can be found following the same procedure, but it doesn't admit a closed form, unless $V_a=V_b=V_0$ (in that case the population size $N$ is just rescaled by a factor $(1+V_0)$).  
The analysis of the fixed points of the deterministic system $y^*$ and of the extrema of the p.d.f. $\tilde y$ can still be carried out, starting from the two equations
\begin{equation}
    F(y^*)=0,\quad F(y) = N \left\{\frac{s}{1+sy}\left(\frac{1}{4}-y^2\right)-\frac{2\nu y}{1+sy}\right\}-\frac{w}{(1+sy)^2}\left(\frac{1}{4}-y^2\right);
    \label{fp-mb}
\end{equation}
\begin{equation}
    F(\tilde y) - \frac{1}{2}\partial_y D(\tilde y)= 0, \quad D(y) = \left(\frac{1}{4}-y^2\right)\left[1+\frac{V_0 + w y}{(1+sy)^2}-\frac{2wy}{(1+sy)^2}\left(\frac{1}{2}+y\right)\right]. 
    \label{ext-mb}
\end{equation}
In contrast to the previous case, $y^*$ and $\tilde y$ will now generally depend not only on $\nu$ and $s$, but also on $w=(V_a-V_b)/V_0$ and $V_0=(V_a+V_b)/2$. A summary of the analysis of the noise-induced transition for the two-species population dynamics is plotted in Fig~\ref{fig:bifurc-diags}.

\section{Correction to Kimura's formula and impact of mini-batching in the equilibrium regime}%Correction to Kimura's formula and free fitness} 
The implicit regularization that emerges from environment mini-batching is fundamentally a non-equilibrium phenomenon. 
Nonetheless, when evolution is very slow, such that it proceeds via a sequence of fixation or extinction events of very rare, randomly appearing mutations in isogenic populations, the population dynamics can be cast into the framework of equilibrium statistical mechanics. 
Sella and Hirsch have shown that the steady state distribution of such a process can be described in terms of a (negative) thermodynamic potential dubbed \emph{free fitness} \cite{Sella2005}. 
We can analyze our system in this tractable regime and study how the introduction of environment mini-batching affects the functional form of the free fitness potential. 

The core of the calculation consists in gathering how mini-batching modifies Kimura's orginal formula, which describes the fixation probability of a nearly neutral mutation in a wild type population. This is indeed the irreversible event where the non-equilibrium character of population dynamics manifests in the theory, and where the annealed population heterogeneity can hence play a role.

The derivation of Kimura's formula \cite{Kimura-original} relies on the diffusion approximation limit, from which a backward Kolmogorov equation for the first passage probability can be obtained:
\begin{equation}
    \partial_t u (z,t) = -F(z) \partial_z u(z,t) + D(z) \partial_z^2 u(z,t)
    \label{Kolmogorov-bkd}
\end{equation}
where $u(z,t)$ is the probability of being fixed by time $t$, starting from a fraction $z$ of the mutant in the population. The associated boundary conditions are: $u(0,t)=0$, $u(1,t)=1$. Define $u(z) = \lim_{t\to\infty}u(z,t)$: this is the solution of the stationary problem \eqref{Kolmogorov-bkd}. Since the system is one-dimensional, there exists an easy formal solution:
\begin{equation}
    u(z) = \frac{\int_0^z g(z') dz'}{\int_0^1 g(z') dz'}, \qquad g(z)= e^{\int_{z_0}^z{ \frac{2 F(z')}{D(z')}dz'} }.
\end{equation}
Without loss of generality, let us take the mid fitness of the two quasi-species (wild type and mutant) to be $f_0=1$, and $V_0$ the mid variance of the two species (in units of $f_0^2$). Let us finally rename $s=f_{\mathrm{mut}}-f_{\mathrm{wt}}$ and $w=V_{\mathrm{mut}}-V_{\mathrm{wt}}$ the effects of the mutation on mean and variance of the fitness. We can rewrite the drift and diffusion terms of the one-dimensional Ito-SDE as a function of these parameters:
\begin{equation}
    F(z)=\frac{Nsz(1-z)}{1+s(z-1/2)}- \frac{wz(1-z)}{[1+s(z-1/2)]^2}\approx (Ns-w)z(1-z) +O(s^2,sw);
    \label{Taylor-F}
\end{equation}
\begin{equation}
    D(z)= z(1-z)\left(1+\frac{V_0 - w(z-1/2)}{\left[1+s(z-1/2)\right]^2}\right) \approx z(1-z) (1+ V_0) +O(s,w)
    \label{Taylor-D}
\end{equation}
where the approximations come from the assumption that $s\ll 1$ and $w\ll 1$. 
In this limit, the solution is very simple:
\begin{equation}
    u(z)\approx\frac{1-e^{\frac{2(Ns-w)}{1+V_0}z}}{1-e^{\frac{2(Ns-w)}{1+V_0}}}.
\end{equation}
and we obtain a simple modification to Kimura's formula:
\begin{equation}
    K(s,w) = u(1/N) = \frac{1-e^{\frac{2(Ns-w)}{N(1+V_0)}}}{1-e^{\frac{2(Ns-w)}{1+V_0}}}.
\end{equation}
Notice that in addition to the fitness advantage $s$ of the mutant, Kimura's formula becomes now dependent on the mid fitness variance $V_0$ between the mutant and wild type, and on the difference between their fitness variances $w$.

Following the line of reasoning of Sella and Hirsch \cite{Sella2005}, we restrict to a space of population states of uniform genetic composition, whose steady-state probabilities are indicated by $\{\pi_a\}$. Transition rates between these states are indicated by $W_{ab} = \mu_{ab}K(s,w)$, if $s=(f_a-f_b)$, $w = (V_a - V_b)/f_0^2$, and if $\mu_{ab}$ is the (small) rate at which mutation $a$ emerges from $b$. At the same order of the Taylor expansion used in \eqref{Taylor-F}--\eqref{Taylor-D}, we can reinterpret $f_0$ and $V_0$ as the average mean fitness and the average fitness variance across the whole genotypic space.
Detailed balance gives:
\begin{equation}
    \frac{W_{ab}}{W_{ba}}=\frac{K(s,w)}{K(-s,-w)}=e^{\frac{2(Ns-w)}{N(1+V_0)}(1-N)}=\frac{\pi_a}{\pi_b} = e^{\beta(\varepsilon_b-\varepsilon_a)}\implies \varepsilon_a = -f_a + \frac{V_a}{N},\quad \beta = \frac{2(N-1)}{1+V_0}.
\end{equation}

\begin{figure}
    \centering
    \includegraphics[width=0.95\linewidth,trim={0 0 0 6cm},clip]{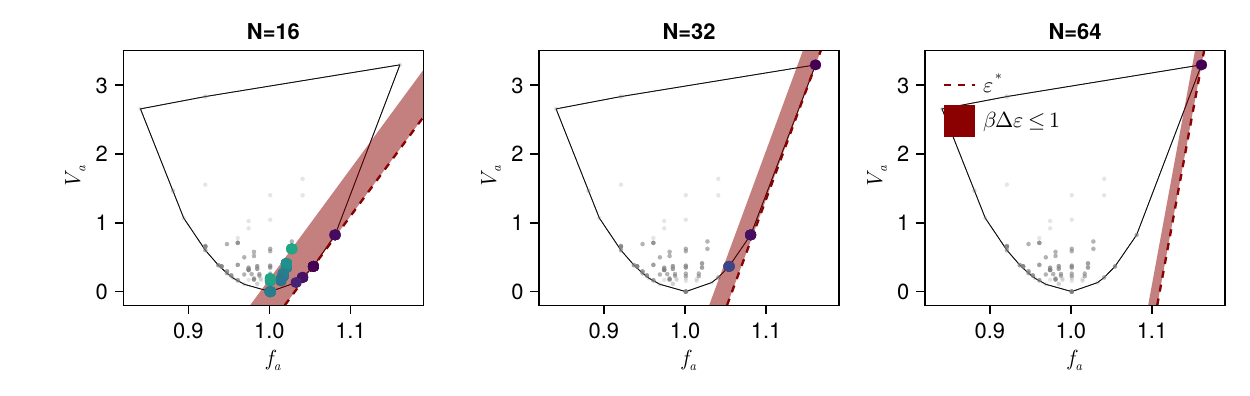}
    \caption{\textbf{Equilibrium populations.} %\emph{First row-} We represent,a s a function of $N$, the lowest energy genotype (solid line) and the states whose energy value is within $\beta^{-1}$ distance from that (red points).  \emph{Second row-} 
    Representation in the $f-V$ plane of the states that would be significantly populated at equilibrium, for different values of $N$. The color of the dots indicates the value of $\Delta\varepsilon$. }
    \label{fig:eq-pop}
\end{figure}

Having the Boltzmann distribution $\pi_a=Z^{-1}e^{-\beta \varepsilon_a}$, we can construct any thermodynamic potential of interest. The most straightforward one is the free energy analogue, $G=\beta^{-1}\ln Z$, i.e. the negative of the free fitness introduced in \cite{Sella2005}. Its expression reads
\begin{equation}
    G = \langle \varepsilon\rangle - \beta^{-1}S[\pi] = -\langle f\rangle + \frac{\langle V\rangle}{N} + \frac{1+V_0}{2(N-1)}\langle\ln \pi\rangle.
\end{equation}
In analogy to \cite{Sella2005}, it can be shown that $G$ is a Lyapunov function for the dynamics of the Markov chain, and its minimization determines the equilibrium state of the system. As pointed out in \cite{Sella2005}, this minimization principle reduces to Fisher's fundamental theorem of evolution in the limit of infinite population size; when the population is finite, deviations arise from the so-called mutational load, quantified by the entropic term $S[\pi]$, as well as from $\langle V\rangle/N$, a term coming from the correction to the deterministic selection force introduced by single-cell variability.

We can also use the Boltzmann distribution to identify which genotypes are populated at each population size $N$. 
In the $N\to\infty$ limit, we shall expect the population to be uniquely concentrated on the fittest genotype, as minimizing $\varepsilon$ corresponds to maximizing $f$. At finite population sizes, we can match the energy gap with the scale of thermal fluctuations (cfr. Fig~\ref{fig:eq-pop}) to identify the critical population size at which the probability is no longer dominantly concentrated only on the fittest genotype:
\begin{equation}
    \beta_N|\varepsilon_N^0- \varepsilon_N^i|\geq1 \forall i \implies N\geq N_c = \max_{(s,w)}\left(\frac{1+V_0}{2s}+\frac{w}{s}+1\right)\approx \max_s\frac{1+V_0}{2s}
    \label{eq:Nc}
\end{equation}
where we have neglected higher orders in $s$ and $w$. From a comparison with classical results \cite{Sella2005}, this formula highlights again that, at leading order, the effect of individual variability is akin to shrinking the population size by a factor $1+V_0$, where $V_0$ is the reference level of cell-to-cell variability for the species involved.

\begin{figure}[h]
    \centering
    \includegraphics[width=0.95\linewidth]{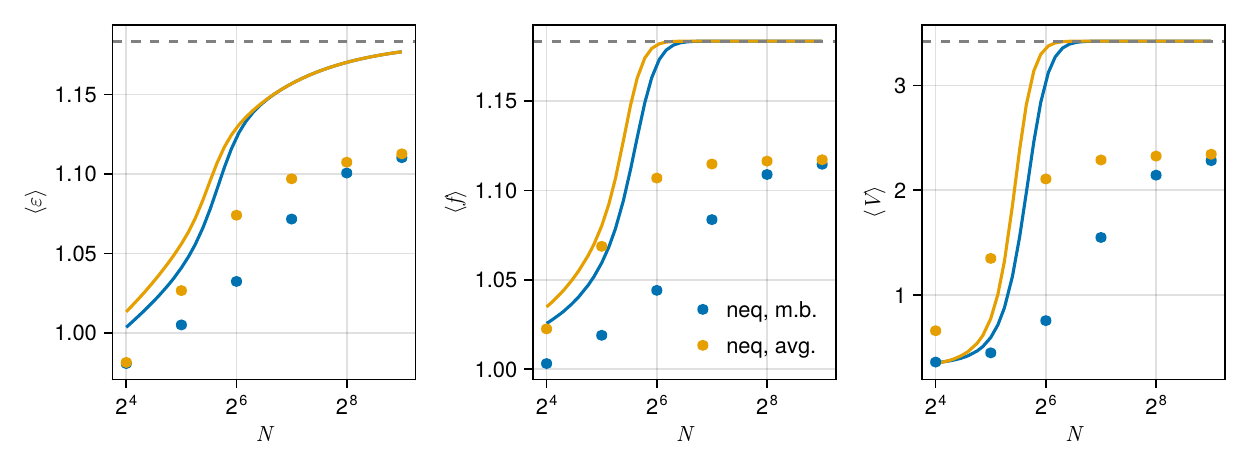}
    \caption{\textbf{Equilibrium vs non-equilibrium steady states.} Comparison of the average values of the three observables $\varepsilon = f - V/N$ (Boltzmann energy), $f$ (fitness mean) and $V$ (fitness variance) in the equilibrium and non-equilibrium steady states. These averages are computed from the analytical formulas in the equilibrium-like regime, and from numerical simulations out of equilibrium (with and without environment mini-batching).}
    \label{fig:eq-vs-noneq}
\end{figure}

We remark that, in order to make analytical progress, we exploited here the approximation of near-uniform variability, $w\ll 1$. This approximation inherently cannot capture the magnitude of the thermophoretic effect observed in Fig. 2. The discrepancy between the numerical results in the non-equilibrium regime and the theoretical predictions at equilibrium (shown in Fig.~\ref{fig:eq-vs-noneq})
is not only due to the limitations of the small-$w$ approximation, but also to the omitted impact of mutations on the population dynamics in the presence of individual variability.
At elevated mutation rates, where the equilibrium treatment breaks down, the long-lived states are no longer pure states, but more mixed population states, where the demographic noise and its corrections matter most. Despite these limitations, this simple analysis offers valuable insights into how stochasticity enhances fitness landscape navigability, which can be naturally incorporated into high-throughput phenotyping studies: the equilibrium framework indeed continues to serve as the understood foundation for interpreting most phenotypic landscape reconstructions.

%%% Each figure should be on its own page

\begin{figure}
    \centering
    \includegraphics[width=0.95\linewidth]{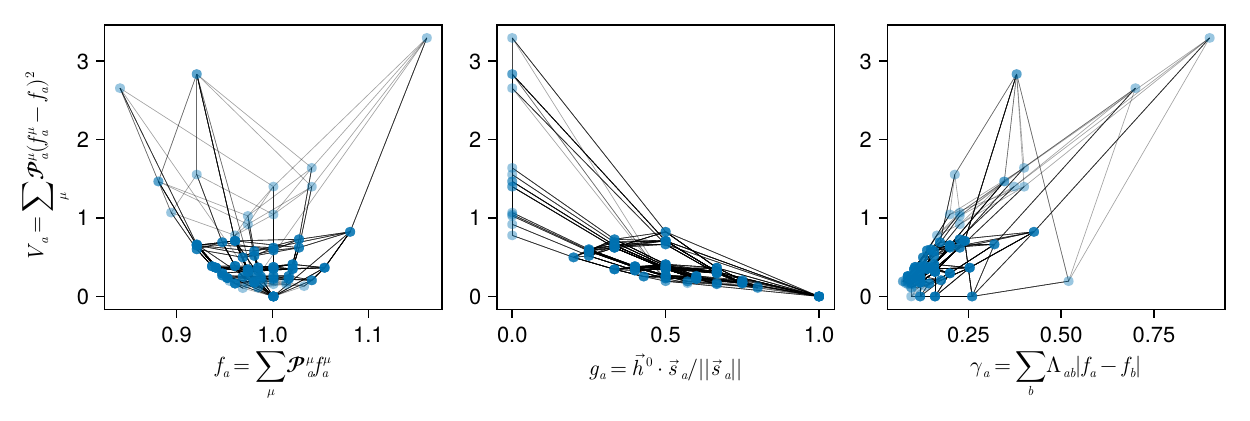}
    \caption{\textbf{Design of the landscape ensemble.} Constitutive relation between mean and variance (\emph{left}), generalization and variance (\emph{center}) and inverse flatness and variance (\emph{right}) for the fitness landscape ensemble used for the analyzed example in Fig. 2 and Fig. 3 in the main text}
    \label{fig:features}
\end{figure}

\begin{figure}
    \centering
    \includegraphics[width=\linewidth]{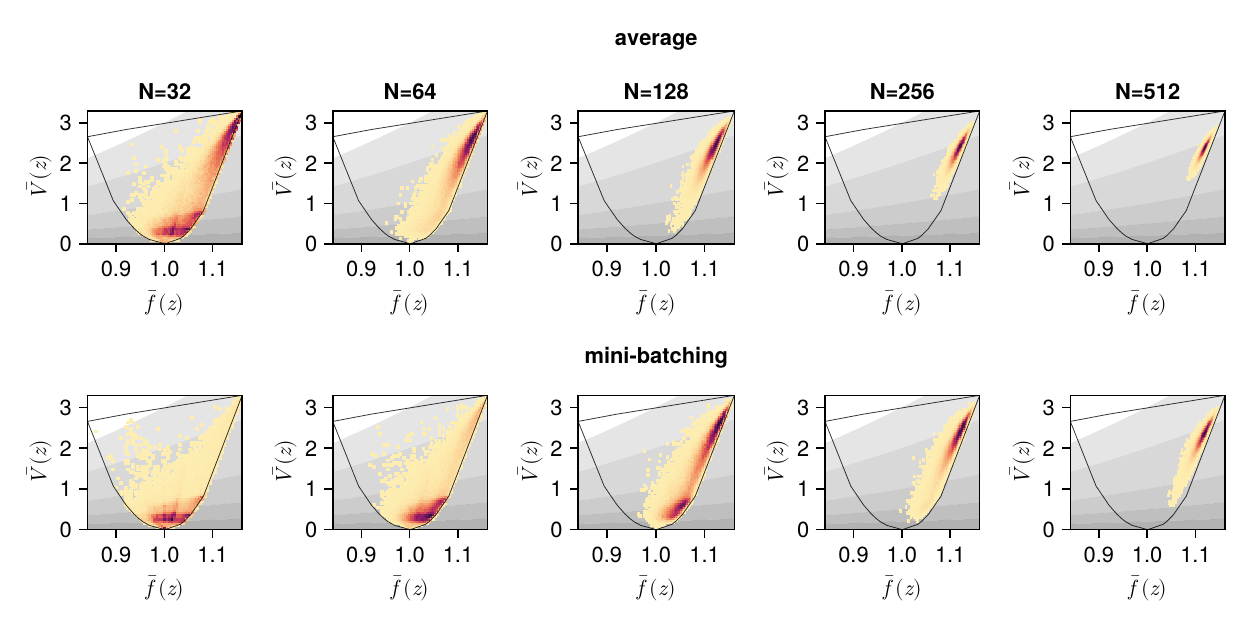}
    \caption{\textbf{Impact of mini-batching on steady state distributions of population states}. We empirically reconstruct the joint probability distribution $P\left(\bar f(\boldsymbol z),\bar V(\boldsymbol z)\right)$ at different values of $N$. In general, the center of mass is shifted towards lower values of average variance, i.e. better generalization score. The shaded level sets correspond to the approximate rescaling factor of the demographic noise term or effective temperature, $1+ \bar V/\bar f^2$.}.
    \label{fig:S10}
\end{figure}

\begin{figure}
    \centering
    \includegraphics[width=\linewidth]{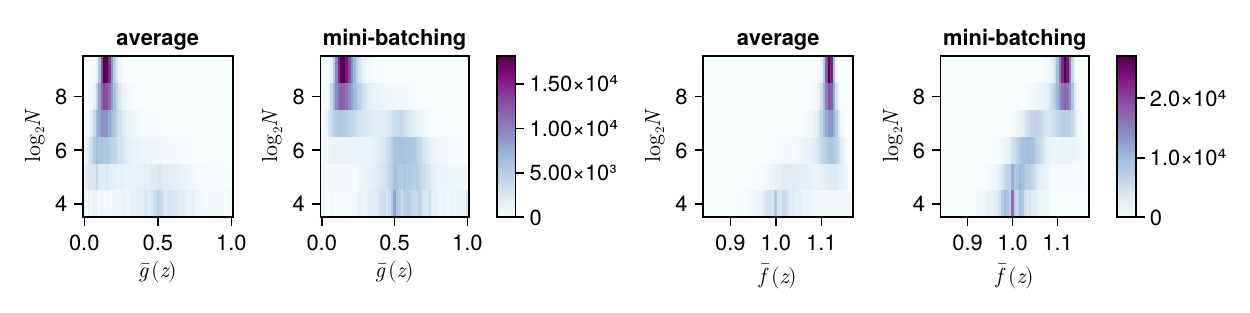}
    \caption{\textbf{Noise-induced phase transition.} Comparison of the empirical distributions of the collective quantities $\bar g(\boldsymbol z)$ (\emph{left}) $\bar f(\boldsymbol z)$ (\emph{right}) obtained from numerical simulations of the population dynamics with and without mini-batching in the fitness landscape described in Fig. 2. In the presence of mini-batching, the transition point is shifted to larger values of $N$, and a clearer bimodality is observed at the transition.}
    \label{fig:S9}
\end{figure}

\begin{figure}
    \centering
    \includegraphics[width=0.9\textwidth]{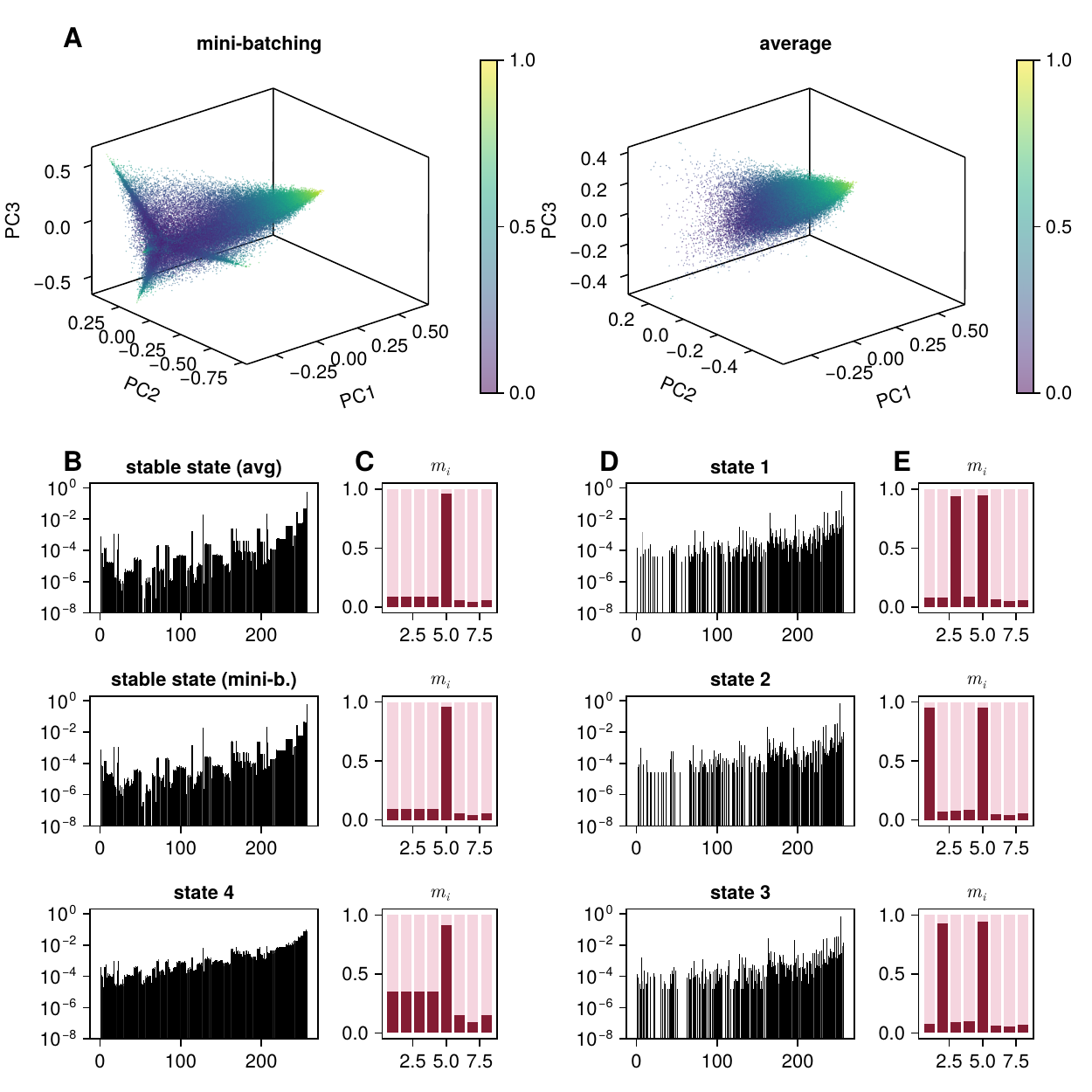}
    \caption{\textbf{Characterization of population states.} \textbf{A PCA of the steady states.} We project the steady states of the population ($N=128$) along the first three principal components of the data obtained from 100 trajectories of the process with mini-batching at stationarity (126 points per trajectory). The total number of trajectories represented in each scatterplot is 700. The first PC explains 50\% of the variance and separates the steady state occupied in the reference process from the aggregate of population states defined as the meta-stable state in the main text. The PC representation of the metastable state highlights the additional internal structure, with a core "mixed state" of low Inverse Participation Ratio (IPR) (dark color) and 3 main "pure sates" with high IPR (light color). The IPR of the population sate is computed as follows: $IPR = \sum_a z_a^2$. \textbf{B Characterization of the stable and metastable states.} We plot the mean composition of the population in the stable state of the reference process (first row,}
\end{figure}

\begin{figure}[h]
\ContinuedFloat
\caption{from all the data in the right panel of A), in the stable state of the mini-batching process (second row, from the points with PC1$>$0), and in the metastable state (third row, from all the points with PC1$\leq$0 in the left panel of A). The mean stable states are nearly identical, and they exhibit strong localization on the species of highest fitness, accounting alone for 60\% of the population. The metastable state resembles more a mixed state, on average. \textbf{C Conservation profiles of the stable and metastable states.} Each individual in the population is associated to a binary sequence,. Given the mean population composition in each state, we compute the mean binary sequence associated to them, $\vec m$, with $m_i=\sum_as_i^a\langle z_a\rangle_{state}$. The mean sequence profile of the metastable state confirms strong conservation of the fittest genotype, with $m_i$ values consistently close to 0 or 1. The metastable state exhibits good conservation on the sites where the selection fields vary ($K < i \leq L$) and low conservation on the sites where the selection fields are constant ($1\leq i \leq K$). \textbf{D Sub-characterization of the metastable state.} The metastable state is organized along three main axes that emanate from the center of the PC2-PC3 plan and terminate on the vertices of a triangle, where the IPR peaks. We incorporate into State 1, State 2 and State 3, respectively, the sets of points that fall at a square distance less than 0.05 from any of the three vertices. There is string localization on high fitness genotypes, which are nonetheless different from the highest-fitness genotype (genotypes are ordered by increasing fitness). \textbf{E Conservation profiles of the metastable sub-states.} The typical sequence sampled from each of the metastable sub-states has exactly two nonzero entries: $s_5=1$ and $s_k=1$ for some $k$ among the first $K$ sites subject to constant selection field. Hence the generalization score of the typical sequences is 0.5, and their average fitness 1.08.}
\end{figure}

\begin{figure}
    \centering
    \includegraphics[width=0.85\linewidth]{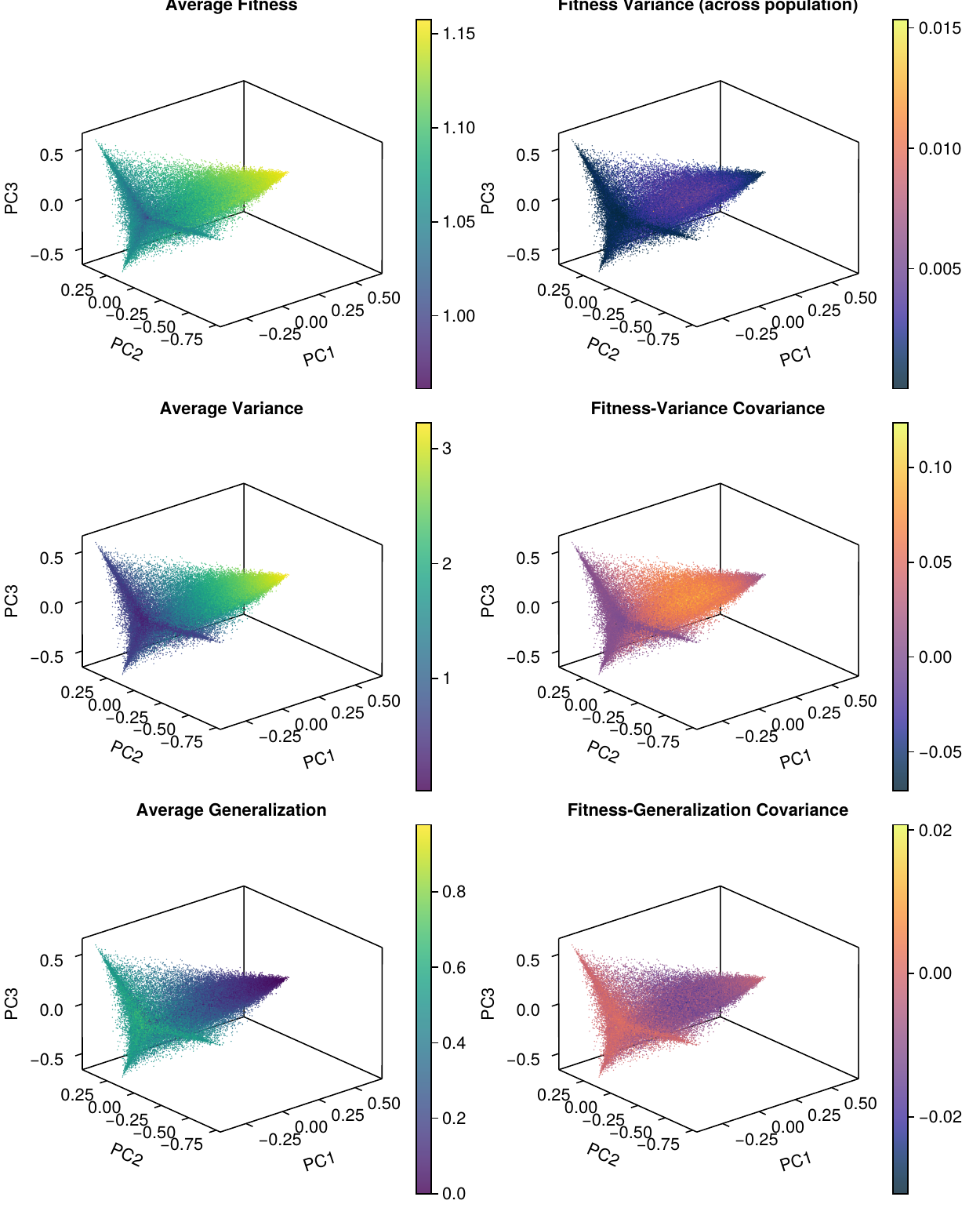}
    \caption{\textbf{Association of PCs and observables.} Left column: first moments (means) of the indicated observables across the population distributions $\boldsymbol z$, i.e. $\bar A  = \sum_a A_a z_a$. Right column: second moments (variances, covariances) of the indicated variables across the population distribution, i.e $Cov_z(A,B) = \sum_a A_aB_az_a - \bar A\bar B$. The variables in the right column are the main determinants of the deterministic drift in the Price equation for the corresponding observable on the left column.}
    \label{fig:enter-label}
\end{figure}

\newpage
\bibliography{apssamp}% Produces the bibliography via BibTeX.

\end{document}